\documentclass[a4paper, amsfonts, amssymb, amsmath, reprint, showkeys, footinbib, twoside,superscriptaddress,floatfix,longbibliography]{revtex4-1}

\usepackage{graphicx} % Required for inserting images
\usepackage{amsmath}
\usepackage{comment}
\usepackage{booktabs}
\usepackage{hyperref}
\usepackage{url}
\usepackage{adjustbox}
\usepackage{array}
\usepackage{babel}

\usepackage{multirow}
\usepackage{hhline}
\usepackage{makecell}
\usepackage[table]{xcolor}

\newcommand{\figLabelCapt}[1]{\textbf{\MakeLowercase{{#1}}}}
\newcommand{\refSub}[2]{\hyperref[#2]{\ref{#2}\figLabelCapt{#1}}}
\newcommand{\figref}[1]{Fig.~\ref{#1}}

\newcommand{\br}[1]{\mathbf{r}}
\newcommand{\bk}[1]{\mathbf{k}}

\begin{document}

\title{A universal augmentation framework for long-range electrostatics \\in machine learning interatomic potentials}

\author{Dongjin Kim}
\affiliation{Department of Chemistry, UC Berkeley, California 94720, United States}

\author{Xiaoyu Wang}
\affiliation{Department of Chemistry, UC Berkeley, California 94720, United States}

\author{Peichen Zhong}
\affiliation{Bakar Institute of Digital Materials for the Planet, UC Berkeley, California 94720, United States}

\author{Daniel S. King}
\affiliation{Bakar Institute of Digital Materials for the Planet, UC Berkeley, California 94720, United States}

\author{Theo Jaffrelot Inizan}
\affiliation{Bakar Institute of Digital Materials for the Planet, UC Berkeley, California 94720, United States}

\author{Bingqing Cheng}
\email{bingqingcheng@berkeley.edu}
\affiliation{Department of Chemistry, UC Berkeley, California 94720, United States}
\affiliation{Bakar Institute of Digital Materials for the Planet, UC Berkeley, California 94720, United States}
\affiliation{The Institute of Science and Technology Austria, Am Campus 1, 3400 Klosterneuburg, Austria}

\date{\today}

\begin{abstract}
Most current machine learning interatomic potentials (MLIPs) rely on short-range approximations, without explicit treatment of long-range electrostatics. To address this, we recently developed the Latent Ewald Summation (LES) method, which infers electrostatic interactions, polarization, and Born effective charges (BECs), just by learning from energy and force training data. Here, we present LES as a standalone library, compatible with any short-range MLIP, and demonstrate its integration with methods such as MACE, NequIP, CACE, and CHGNet. We benchmark LES-enhanced models on distinct systems, including bulk water, polar dipeptides, and gold dimer adsorption on defective substrates, and show that LES not only captures correct electrostatics but also improves accuracy. Additionally, we scale LES to large and chemically diverse data by training MACELES-OFF on the SPICE set containing molecules and clusters, making a universal MLIP with electrostatics for organic systems including biomolecules. MACELES-OFF is more accurate than its short-range counterpart (MACE-OFF) trained on the same dataset, predicts dipoles and BECs reliably, and has better descriptions of bulk liquids. By enabling efficient long-range electrostatics without directly training on electrical properties, LES paves the way for electrostatic foundation MLIPs.
\end{abstract}

\maketitle

\section{Introduction}
The short-range approximation underlies
most established machine learning interatomic potentials (MLIPs)~\cite{keith2021combining,unke2021machine},
as it enables the decomposition of total energy into individual contributions from local atomic environments,
allowing efficient learning and inference.
However, it is increasingly recognized that explicitly modeling long-range interactions is essential for systems with significant electrostatics and dielectric response, 
such as electrochemical interfaces~\cite{niblett2021learning}, charged molecules~\cite{grisafi2019incorporating,huguenin2023physics}, and ionic~\cite{zhang2022deep} and polar materials~\cite{monacelli2024electrostatic}.

While various approaches have been developed to incorporate long-range interactions into MLIPs~\cite{unke2019physnet,ko2021fourth,gao2022self,sifain2018discovering,gong_predictive_2025,shaidu2024incorporating,yu2022capturing,kosmala2023ewald,grisafi2019incorporating,huguenin2023physics,ple_force-field-enhanced_2023,inizan_scalable_2023,monacelli2024electrostatic,caruso2025Extending}, many require specialized training labels beyond energy and forces: the fourth-generation high-dimensional neural network potentials (4G-HDNNPs)~\cite{ko2021fourth}, the 
charge constraint ACE model~\cite{Rinaldi2025Chargeconstrained}, and
BAMBOO~\cite{gong_predictive_2025}
are trained to reproduce atomic partial charges from reference quantum mechanical calculations;  the self-consistent field neural network (SCFNN)~\cite{gao2022self} and the deep potential long-range (DPLR) model~\cite{zhang2022deep} learn from the positions of maximally localized Wannier centers (MLWCs) for insulating systems; and 
PhysNet~\cite{unke2019physnet}, AIMNET2~\cite{Anstine2025}, and SO3LR~\cite{Kabylda2025} utilize the dipole moments of gas-phase molecules.
These additional data requirements limit the applicability of such methods to standard datasets containing only atomic positions, energies, forces, and sometimes stresses.

By contrast, a few methods can learn long-range effects directly from standard datasets~\cite{yu2022capturing,kosmala2023ewald,grisafi2019incorporating,huguenin2023physics,Faller2024Densitybased,monacelli2024electrostatic}. These include message-passing schemes such as Ewald message passing~\cite{kosmala2023ewald} and RANGE~\cite{caruso2025Extending}, and Coulomb-inspired long-range message-passing model FeNNix~\cite{ple2025Foundation}, and descriptor-based approaches like LODE~\cite{grisafi2019incorporating,huguenin2023physics,Loche2025Fastb} and the density-based long-range descriptor~\cite{Faller2024Densitybased}. However, in these methods the long-range contributions are not explicitly related to electrostatics due to charged atoms,
making it difficult to extract electrical response properties.

The Latent Ewald Summation (LES) framework~\cite{Cheng2025Latent,kim2024learning,zhong2025machine} overcomes these limitations by inferring atomic charges and the resulting long-range electrostatics directly from total energy and force data, without training on explicit charge labels. 
LES can thus be trained on any dataset used for short-ranged MLIPs.
The inference of atomic charges not only enhances physical interpretability, but also enables the extraction of polarization ($P$) and Born effective charges (BECs)~\cite{zhong2025machine}. 
These quantities allow computation of electrical response properties, including dielectric constants, ionic conductivities, ferroelectricity, and infrared (IR) spectra, 
and also enable MLIP-based molecular dynamics under applied electric fields.

Here, we introduce LES as a universal electrostatics augmentation framework for short-ranged MLIPs. 
Unlike prior methods that are tightly integrated with specific architectures, LES is implemented as a standalone library that can be combined with a wide range of MLIP models~\cite{behler2007generalized, bartok2010gaussian, shapeev2016moment, drautz2019atomic, batzner20223, batatia2022mace, cheng2024cartesian, qu2024the, ko2025_matgl, fu2025Learning, pozdnyakov2023smoothE, Pelaez2024TorchMDNeta, haghighatlari2022newtonnet, liao2023Equiformer}. 
We provide open-source patches for several MLIP packages (MACE~\cite{batatia2022mace}, NequIP~\cite{batzner20223}, CACE~\cite{cheng2024cartesian}, MatGL~\cite{ko2025_matgl}), enabling seamless integration. 
We benchmark and compare LES-augmented models across diverse systems with the baseline short-ranged MLIPs, demonstrating improved accuracy and correct long-range electrostatics through the inference of BECs.

Additionally, to demonstrate scalability to large, chemically diverse datasets, we train MACELES-OFF, an MLIP with explicit long-range electrostatics for organic molecules, using the SPICE dataset~\cite{eastman2023spice}. We find that MACELES-OFF outperforms its short-range counterpart (MACE-OFF~\cite{kovacs2025MACEOFF}) and accurately predicts BECs, despite never being trained on charge or polarization data.

\section{Architecture}

\begin{figure*}[t]
\centering
\includegraphics[width=0.7\linewidth]{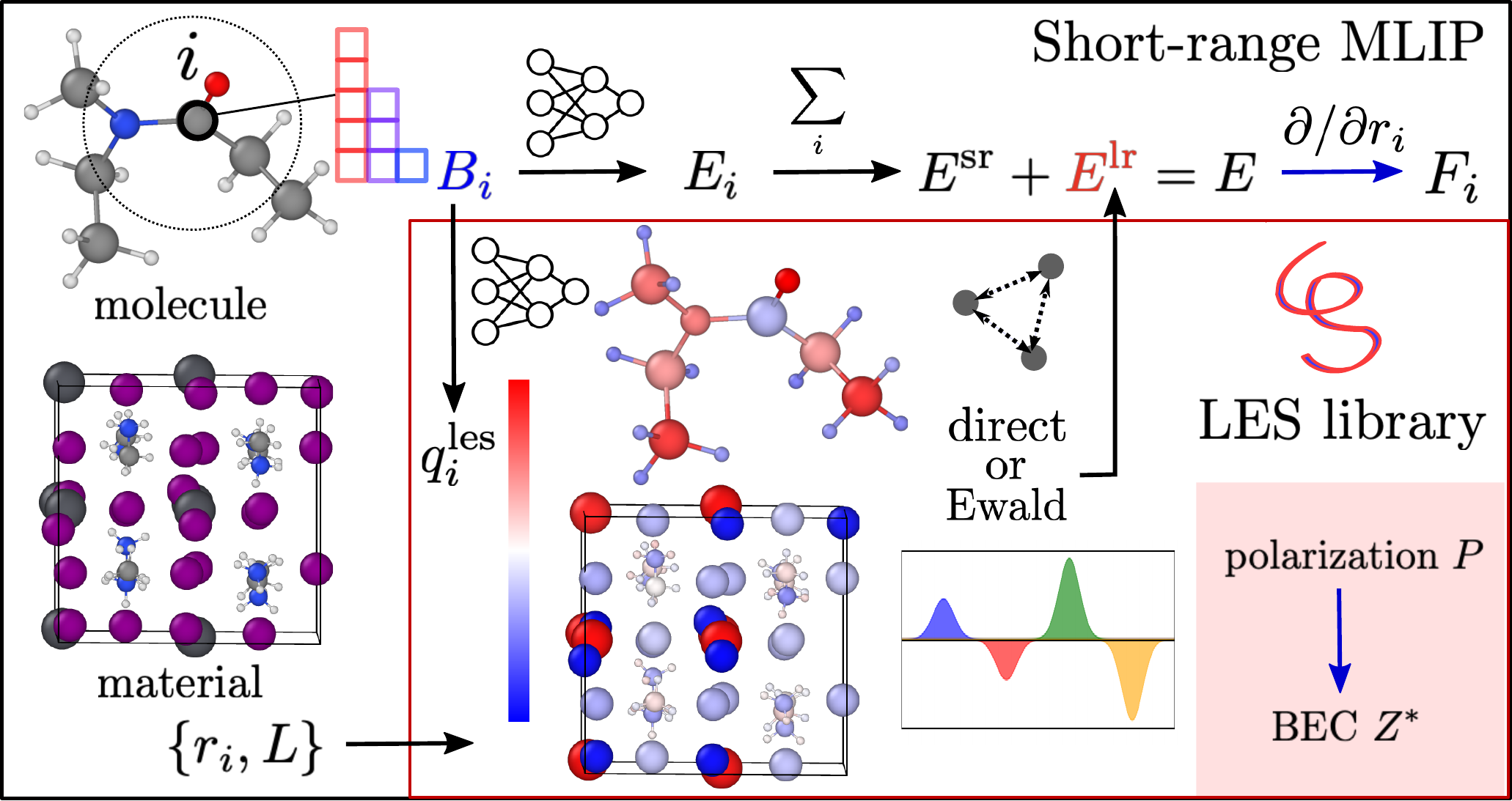}
     \caption{
Schematic illustration of the LES integration with a short-ranged MLIP.  
The black box shows a standard MLIP workflow, where invariant or equivariant atomic features are computed, the invariant features $B_i$ are mapped to atomic energies $E_i$ and summed to yield the short-range energy $E^{\mathrm{sr}}$.
The red box shows the LES module, which either predicts latent charges $q_i^\mathrm{les}$ from atomic features $B_i$ or receives them from the host MLIP. 
Based on atomic positions $r_i$ and simulation cell size $L$, LES then computes the long-range energy $E^{\mathrm{lr}}$ via Ewald summation or pairwise interactions.
Optionally, LES also enables the computation of Born effective charges (BECs) as shown in the shaded red region.
The blue arrows indicate auto-differentiation operations
that are used to compute forces $F_i$ or BECs.
    }
    \label{fig:les-schematic}
\end{figure*}

The integration of the LES library with a host MLIP is illustrated in Fig.~\ref{fig:les-schematic}.
The black box outlines the standard workflow of a short-range MLIP: 
Given a molecular or periodic system, the base MLIP computes invariant and/or equivariant features for each atom $i$, based on either local atomic environment descriptors~\cite{drautz2019atomic,behler2007generalized,wang2018deepmd} or message-passing architectures~\cite{schutt2017schnet}.
The local invariant descriptors $B_i$ are mapped via a neural network to atomic energies $E_i$, which are then summed to obtain the total short-range energy
$E^{\mathrm{sr}} = \sum_{i=1}^N E_i$.

The red box in Fig.~\ref{fig:les-schematic} shows the LES module that is written in PyTorch, and the black arrows crossing the two boxes indicate communication between LES and the host MLIP.
The LES module can either take the
 $B_i$ features to predict the latent charge of $i$ of each atom, $q^\mathrm{les}_i$, via a neural network,
 or this charge prediction part can happen inside the MLIP and be passed to LES.
These $q^\mathrm{les}$ charges are then used to compute
the long-range energy contribution $E^\mathrm{lr}$. 
For periodic systems, the Ewald summation is used:
\begin{equation}
E^\mathrm{lr} = \dfrac{1}{2\varepsilon_0V} \sum_{0<k<k_c} \dfrac{1}{k^2} e^{-\sigma^2 k^2/2} |S(\mathbf{k})|^2,
\label{eq:e_lr}
\end{equation}
with
\begin{equation}
S(\mathbf{k}) = \sum_{i=1}^N q_i^\mathrm{les} e^{i\mathbf{k}\cdot\mathbf{r}_i},
\label{eq:sfactor}
\end{equation}
where $\varepsilon_0$ is the vacuum permittivity, $\mathbf{k}$ is the reciprocal wave vector determined by the periodic cell of the system $L$, 
$V$ is the cell volume, and $\sigma$ is a smearing factor (typically chosen to be 1~\AA).
For finite systems, the long-range energy is simply computed using the pairwise direct sum:
\begin{equation}
    E^{\mathrm{lr}}= \frac{1}{2}\frac{1}{4\pi \varepsilon_0} \sum_{i=1}^{N} \sum_{j=1}^N   [1 - \varphi(r_{ij})] \dfrac{q_i q_j}{r_{ij}},
\end{equation}
where the complementary error function $\varphi(r) = \mathrm{erfc}(\frac{r}{\sqrt{2}\sigma})$.

The $E^{\mathrm{lr}}$ computed in LES is returned to the host MLIP package to be added to the total potential energy, $E=E^\mathrm{sr}+E^\mathrm{lr}$.
Inside the host MLIP, forces and stresses are obtained through automatic differentiation of the total energy with respect to atomic positions and cell strain, respectively. 
The training procedure remains unchanged from the original MLIP package, using conventional loss functions for energy, forces, and stresses with user-defined weights. 

Optionally, as indicated by the red shaded area in Fig.~\ref{fig:les-schematic}, LES can predict Born effective
charge tensors ($Z^*_{\alpha\beta}$), where $\alpha$ and $\beta$ indicate Cartesian directions. 
The theory is described in Ref.~\cite{zhong2025machine}.
In brief, for a finite system such as a gas-phase molecule, the BEC for atom $i$ is
\begin{equation}
Z^*_{i \alpha\beta} 
= \dfrac{\partial P_{\alpha}}{\partial r_{i\beta}},
\label{eq:z-finite}
\end{equation}
where $P= \sum_{i=1}^N q_i \mathbf{r}_i$ is the polarization or the dipole moment.
For a homogeneous bulk system under periodic boundary conditions (PBCs), 
\begin{equation}
Z^*_{i \alpha\beta} 
= \Re\left[\exp(-ik r_{i\alpha}) \dfrac{\partial P_{\alpha} (k)}
{\partial r_{i\beta}}\right],
\label{eq:z-pbc}
\end{equation}
with $P_\alpha(k) 
    = \sum_{i=1}^{N} \sqrt{\varepsilon_\infty}\dfrac{q_i^\mathrm{les}}{ik} \exp(i k r_{i\alpha}) $.
The high-frequency (electronic) relative permittivity $\varepsilon_\infty$ is an extra parameter for the system that can be easily obtained from experimental measurements such as refractive index, or from density-functional perturbation theory (DFPT) calculations~\cite{gonze1997dynamical} with frozen nuclei.

Overall, the computation inside LES is light, and the communication between LES and the host MLIP is minimized.
This design allows the LES augmentation to be implemented as a drop-in module that preserves the architecture, training, and inference workflows of the host MLIP package.
As long as the host MLIP is implemented in PyTorch and produces per-atom features used for local energy prediction, LES can be patched in with minimal effort.

Thus far, we have integrated LES into several representative MLIPs with distinct architectures, which we briefly describe here.
NequIP~\cite{batzner20223} is a message passing neural network (MPNN) with E(3)-equivariant convolutions, and it typically uses 3-5 message passing layers with a perceptive field that can be as large as about 20~\AA{}.
CACE~\cite{cheng2024cartesian} is based on atomic cluster expansion (ACE)~\cite{drautz2019atomic}. With one layer ($n_l=1$), CACE is a local atomic descriptor-based model employing polynomial expansions of atomic clusters with body order $\nu$ in Cartesian coordinates.
When $\nu=2$, three-body interactions are captured and (C)ACE has the same expressive power as earlier MLIP methods including HDNNPs~\cite{behler2007generalized}, Gaussian Approximation Potentials~\cite{bartok2010gaussian}, and DeepMD~\cite{zhang2022deep}.
MACE~\cite{batatia2022mace} is a MPNN that combines equivariant representations with the physically grounded structure of atomic cluster expansions. 
Thanks to the high-body-order messages, typically only one message passing layer on top of the base ACE layer is needed.
MatGL is a graph deep learning library for materials modeling \cite{ko2025_matgl}. Instead of using equivariant message passing, MatGL utilizes node and line graph representations to incorporate three-body interactions, such as M3GNet \cite{chen2022_m3gnet} and CHGNet \cite{deng2023chgnet}.
These models span a wide range of architectural paradigms--descriptor-based versus message-passing, invariant versus equivariant representations--and demonstrate that LES can be universally incorporated without requiring architecture-specific modifications.

\section{Benchmarks}

Here, we benchmark the performance of the aforementioned MLIP architectures (MACE, NequIP, CACE, and CHGNet) with and without LES augmentation across three distinct systems, exemplifying the key challenges and practical importance of long-range electrostatic interactions. In the bulk water system, while short-range interactions play an important role in the hydrogen-bond network and structural properties~\cite{remsing2011Deconstructing}, long-range electrostatics are crucial for its dielectric response and interfacial behavior~\cite{gao2020Short}.
The polar dipeptides represent gas-phase systems with strong multipole interactions~\cite{eastman2023spice}.
The Au$_{2}$-MgO(001) system involves the adsorption of metallic species on oxide substrates, where long-range interactions critically affect adsorption energy and charge redistribution, particularly upon substrate doping~\cite{ko2021fourth, kim2024learning}. 
In addition to standard metrics such as energy and force root mean square errors (RMSEs), we also evaluate the prediction of physical observables (BECs and adsorption energies) not in the training set.

\subsection{Water}

\begin{figure*}[t]
\centering
\includegraphics[width=0.95\linewidth]{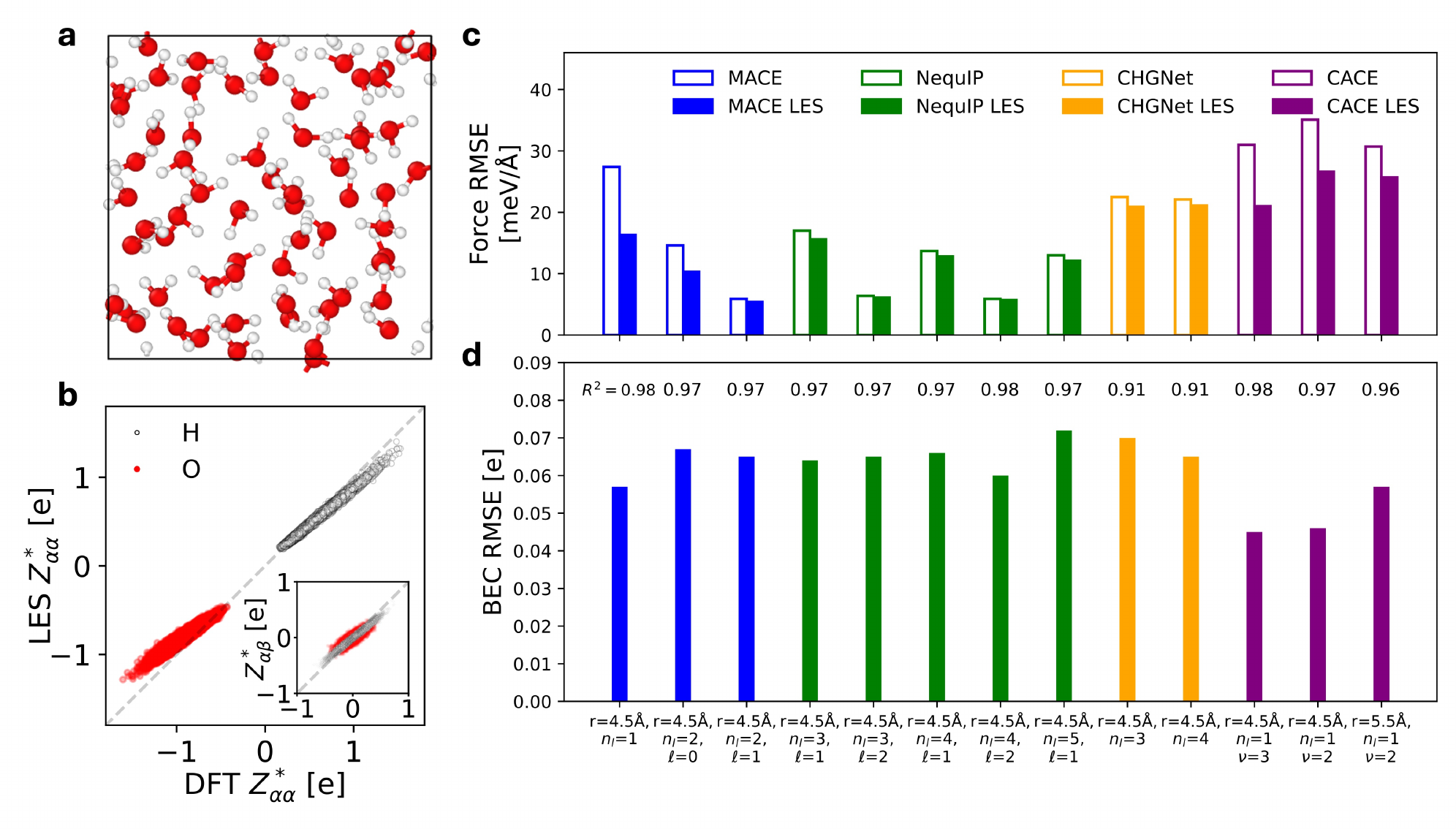}
     \caption{
     Benchmark short-ranged and LES-augmented MLIPs for bulk water.
     \figLabelCapt{a}: A representative snapshot of bulk water.
     \figLabelCapt{b}: A parity plot comparing Born effective charge (BEC) tensors ($Z^*$) for 100 bulk water from LES-augmented NequIP ($r=4.5$~\AA{} model, $n_l$=3, $\ell=2$) against RPBE-D3 DFT~(Ref.~\cite{Schmiedmayer2024}).
     The main panel compares the diagonal elements of BEC ($Z^*_{\alpha\alpha}$), and the inset shows the off-diagonal elements ($Z^*_{\alpha\beta}$ with $\alpha \ne \beta$).
     \figLabelCapt{c}: The force root mean square errors (RMSEs) for baseline MLIPs (hollow bars), and with LES augmentation (solid bars). 
     \figLabelCapt{d}: The RMSEs for BEC predictions using LES and corresponding $R^2$ coefficients.
     The cutoff $r$, the number of layers $n_l$, the order of irreducible representations (rotation order) $\ell$ in message passing, and body order $\nu$ are indicated for each MLIP, as applicable.
    }
    \label{fig:benchmark-water}
\end{figure*}

We trained baseline and LES-augmented MLIPs using the energy and forces from the RPBE-D3 bulk water dataset~\cite{Schmiedmayer2024}, 
which contains 604 train and 50 test configurations, each of 64 water molecules.
%which contains 654 configurations (90\% train/10\% test split) 
%(85.5\% train/4.5\% valid/10\% test split) 
The configurations were generated from MD trajectories at temperatures ranging from 270~K to 420~K at water density under room temperature ($\approx 997$~kg/m$^3$), using an on-the-fly learning scheme. 
For testing LES, we also used the BEC values for an additional 100 snapshots of bulk water at experimental density and room temperature~\cite{Schmiedmayer2024}, computed also from the RPBE-D3 DFT.
We assumed the experimental high-frequency permittivity ($\epsilon_\infty$=1.78) of water in the prediction of BECs based on Eqn.~\eqref{eq:z-pbc}.

Fig.~\ref{fig:benchmark-water}c shows force RMSEs for each MLIP architecture,
as the energy RMSE values remain uniformly low (0.1-0.3~meV/atom) except for CHGNet (0.8-0.9~meV/atom) and are thus omitted here.
For the baseline MLIPs as indicated by the hollow bars in Fig.~\ref{fig:benchmark-water}c, we observe the usual trend typically expected from short-ranged models: 
Without message-passing layers ($n_l=1$), increasing the cutoffs ($r$) improves the accuracy (e.g., comparing CACE $\nu=2$ $r=4.5$~\AA{} and $r=5.5$~\AA{} models), and so does boosting the body order (CACE $r=4.5$~\AA{} $\nu=2$ and $\nu=3$).
More layers ($n_l$) effectively increase the perceptive field of the short-ranged MLIPs and reduce the errors, for both equivariant MPNNs (MACE and NequIP) and the invariant CHGNet.
Increasing the rotation order ($\ell$) in E(3) equivariant representations of messages also helps the accuracy, as seen in both MACE and NequIP models.

The solid bars in Fig.~\ref{fig:benchmark-water}c show the force RMSEs for the LES-augmented models. 
Comparing the solid bars with the hollow ones, LES reduces the errors for all models, while the extent of improvement is somewhat dependent on the architecture of the short-ranged baseline.
For this bulk water dataset, LES benefits most of the MLIPs with a smaller effective perceptive field.
That is, the MACE and the CACE models without message-passing ($n_l=1$) see the largest force error reductions,
and the NequIP and CHGNet with several message-passing layers have modest reductions.
The other hyperparameters of the base MLIPs, $\nu$ and $\ell$, do not seem to have a strong influence on the efficacy of LES.

Moreover, we ask the question: Is LES able to capture the correct electrostatic interactions and predict physical atomic charges?
This is answered via the BEC benchmark shown in Fig.~\ref{fig:benchmark-water}d, which presents the RMSEs of the BECs compared to the DFT calculated values.
Importantly, for all MLIPs, regardless of model architecture, hyperparameters, or effective cutoffs, LES can accurately predict the BECs.
Fig.~\ref{fig:benchmark-water}b shows the parity plot for one of the models, which illustrates the quality of the agreement with the DFT ground truth. 
We do not find any obvious correlation between the accuracy of the BEC prediction and the
baseline MLIP architecture, nor with the force RMSE errors. However, there is a weak inverse correlation with the reduction in force RMSE errors
(see Fig.~\ref{fig:method-force_bec_correlation} in Methods). 
This suggests that LES is a robust approach to infer BECs, regardless of the underlying model performance. 
The BECs are not only sensitive indicators of the correct electrostatics, but are also useful in atomistic simulations for electrical response properties.
For instance, in Ref.~\cite{zhong2025machine}, we showed that such prediction of BECs can enable accurate prediction of water IR spectra under zero or finite external electric fields.

\subsection{Dipeptides}

\begin{figure*}
\centering
\includegraphics[width=0.95\linewidth]{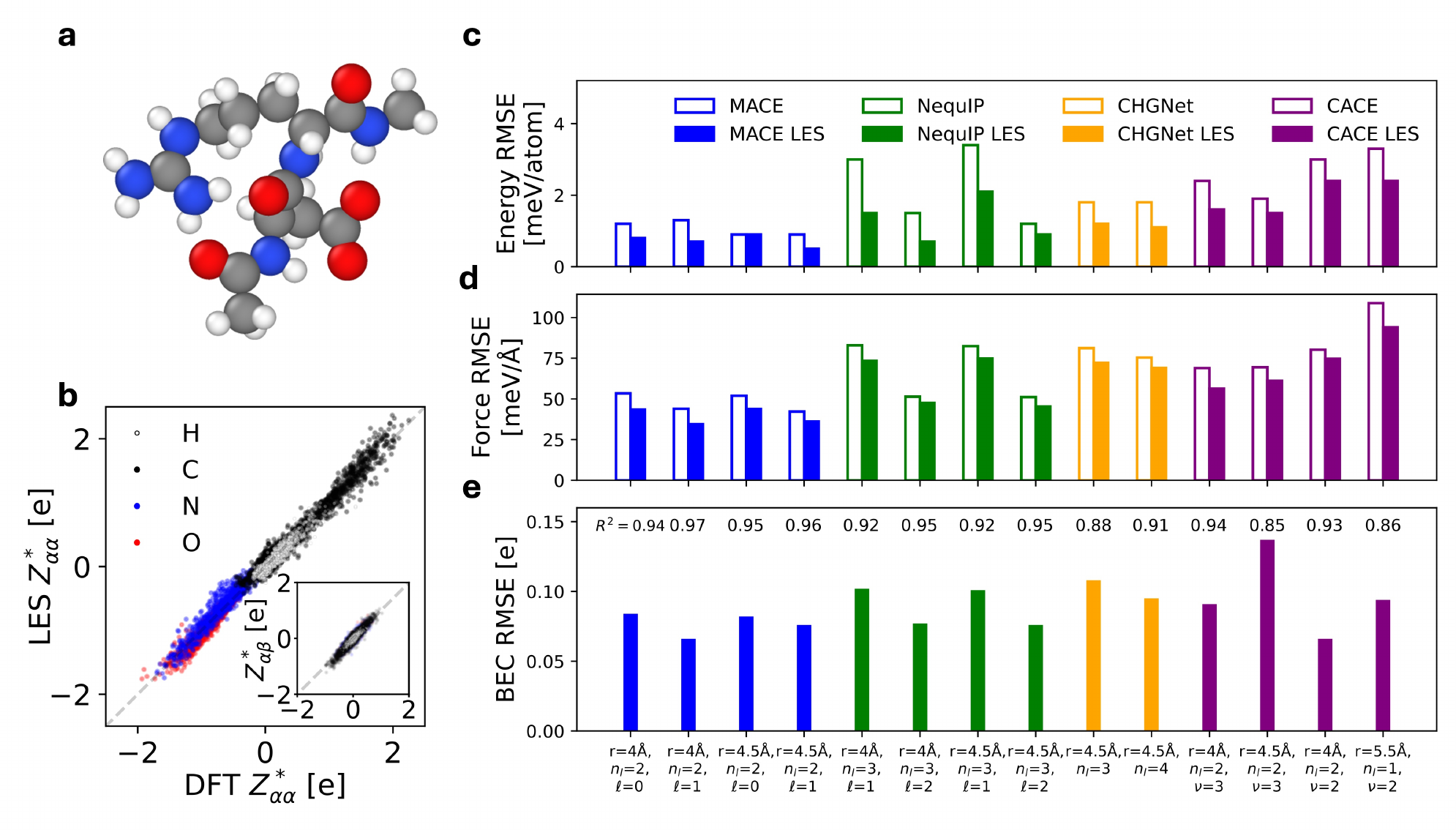}
     \caption{
     Benchmark short-ranged and LES-augmented MLIPs for dipeptide.
     \figLabelCapt{a}: A representative snapshot of a dipeptide conformer composed of arginine and aspartic acid from the SPICE dataset~\cite{eastman2023spice}.
     \figLabelCapt{b}: A parity plot comparing Born effective charge (BEC) tensors for 55 dipeptide configurations~\cite{eastman2023spice} from LES-augmented NequIP ($r=4.5$~\AA{}, $n_l$=3, $\ell=2$) model against DFT. 
     The main panel compares the diagonal elements of BEC ($Z^*_{\alpha\alpha}$), and the inset shows the off-diagonal elements ($Z^*_{\alpha\beta}$ with $\alpha \ne \beta$)     
     \figLabelCapt{c, d}: The energy (c)  and force (d) root mean square errors (RMSEs) for baseline MLIPs (hollow bars), and with LES augmentation (solid bars).
     \figLabelCapt{e}: The RMSEs for BEC predictions using LES and corresponding $R^2$ coefficients.
     The cutoff $r$, the number of layers $n_l$, the order of irreducible representations (rotation order) $\ell$ of messages, and body order $\nu$ are indicated for each MLIP, as applicable.
    }
    \label{fig:benchmark-dipeptide}
\end{figure*}

We trained baseline and LES-augmented MLIPs using energy and forces from a dataset of polar dipeptide conformers selected from SPICE~\cite{eastman2023spice,kim2024learning}, which contains 550 configurations 
(90\% train/10\% test split)
For testing LES, we computed BEC values at the $\omega$B97-mD3(BJ)/Def2SVP level of theory. As these systems are in vacuum, the high-frequency permittivity is $\epsilon_\infty=1$, and one can use Eqn.~\eqref{eq:z-finite} directly when predicting BECs using LES.

Figs.~\ref{fig:benchmark-dipeptide}c and d show the energy and force RMSE values for each MLIP architecture. For the baseline MLIPs indicated by the hollow bars, we again observe the usual trends that increasing body order (CACE $r=4.5$~\AA{} $\nu=2$ and $\nu=3$), and the number of layers (CHGNet $r=4.5$~\AA{} $n_l=3$ and $n_l=4$) lead to improved accuracy. Additionally, increasing $\ell$ improves both MACE and NequIP models.

The solid bars in Figs.~\ref{fig:benchmark-dipeptide}c and d represent the energy and force RMSEs for the LES-augmented models. Comparing the solid and hollow bars, we find again that LES augmentation consistently reduces both energy and force prediction errors across all baseline MLIPs, regardless of their specifics:  
hyperparameters such as $r$ and $\nu$ do not appear to strongly influence the effectiveness of LES, nor do the
perceptive fields of baseline models.
This implies that LES provides an improved molecular representation on top of message passing. 

As is seen in Fig.~\ref{fig:benchmark-dipeptide}e, LES-augmented models all show excellent agreement with DFT reference values for BEC predictions (up to $R^2=0.97$). As an illustration, Fig.~\ref{fig:benchmark-dipeptide}b shows a parity plot.
We also investigated whether the quality of the BEC predictions is correlated with force RMSE values or RMSE reduction, as shown in Figs.~\ref{fig:method-force_bec_correlation}b and d. As is seen, there are no clear-cut correlations between BEC and force or energy RMSE improvement, showing that LES remains a robust way to compute charge regardless of the relative performance of the baseline model. As shown in our previous study, the ability of LES to capture electrostatic interactions also extends to dipole moments, quadrupole moments, and IR spectra~\cite{kim2024learning}. 
Electrostatics strongly affects peptide backbone conformations in dipeptides~\cite{avbelj2006Intrinsic}, which suggests LES augmentation can be helpful for protein structure, folding, binding, and other biological functions. 

\subsection{Au$_{2}$ on MgO(001)} 

\begin{figure*}
\centering
\includegraphics[width=0.95\linewidth]{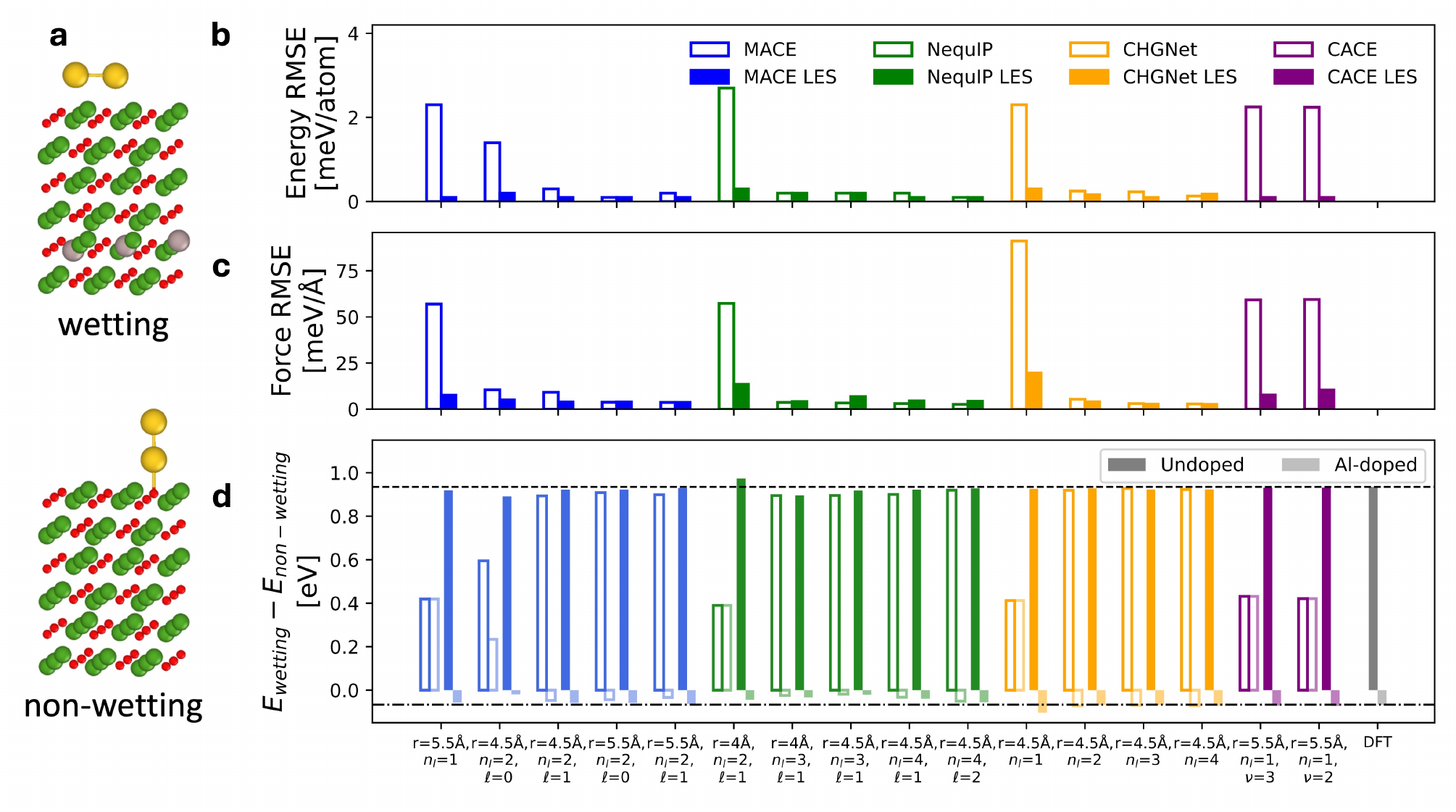}
     \caption{
     Benchmark short-ranged and LES-augmented MLIPs for Au$_{2}$ on MgO(001).
     \figLabelCapt{a}: Representative configurations for wetting (top) or non-wetting (bottom) Au$_{2}$ adsorbed on doped (top) or undoped (bottom) MgO(001) surface.     
     \figLabelCapt{b, c}: The energy (b) and force (c) root mean square errors (RMSEs) for baseline MLIPs (hollow bars), and with LES augmentation (solid bars).
     \figLabelCapt{d}: The energy difference ($E_\mathrm{wetting}-E_\mathrm{non-wetting}$) in eV for doped (top) and undoped (bottom) substrates. Baseline (hollow bars) and LES-augmented models (solid bars) compared with reference DFT results (gray bars and horizontal dashed lines).
     The cutoff $r$, the number of layers $n_l$, the order of irreducible representations (rotation order) $\ell$ of messages, and body order $\nu$ are indicated for each MLIP, as applicable.
    }
    \label{fig:benchmark-Au_MgO}
\end{figure*}

We trained baseline and LES-augmented MLIPs using energy and forces from an Au$_{2}$-MgO(001) dataset from ko et al.~\cite{ko2021fourth}, consisting of 5000 configurations (90\% train/10\% test split). The configurations consist of a gold dimer adsorbed on an MgO(001) surface with two adsorption geometries: an upright non-wetting and a parallel wetting configuration. In some configurations, three Al dopant atoms were introduced in the fifth subsurface layer (Fig.~\ref{fig:benchmark-Au_MgO}a). Despite their distance of over 10~\AA{} away from the gold dimer, the dopant atoms significantly influence the electronic structure and the energetics between the wetting and non-wetting configurations.

Figs.~\ref{fig:benchmark-Au_MgO}b and c present the energy and force RMSEs for each MLIP architecture. 
For all baseline MLIP models indicated by the hollow bars, increasing the perceptive field by either raising the cutoff radius or increasing the number of layers leads to significant improvements in both energy and force accuracy, for both the equivariant MPNNs (MACE, NequIP) and the invariant CHGNet. 
When the perceptive field is smaller than the distance between the dopants and the adsorbates, the MLIPs perform poorly.

The solid bars in Figs.~\ref{fig:benchmark-Au_MgO}b and c demonstrate that the LES augmentation rather consistently improves energy and force predictions across all models, with particularly large gains for models with a smaller perceptive field. For example, the force RMSE reductions for MACE $r=5.5$~\AA{} $n_l=1$, NequIP $r=4$~\AA{} $n_l=2$ $\ell=1$, CHGNet $r=4.5$~\AA{} $n_l=1$, CACE $r=5.5$~\AA{} $n_l=1$ $\nu=3$, and CACE $r=5.5$~\AA{} $n_l=1$ $\nu=2$ are about 80\%. 

Fig.~\ref{fig:benchmark-Au_MgO}d shows the prediction of a physical observable, the energy difference ($E_\mathrm{wetting}-E_\mathrm{non-wetting}$) between wetting and non-wetting configurations for doped and undoped substrates.
This is computed by performing geometry optimizations of the positions of the gold atoms, with the substrate fixed, for both doped and undoped surfaces. 
For the undoped MgO substrate, the non-wetting configuration is more stable, while the Al-doping stabilizes the wetting configuration.
As shown, the baseline MLIPs with small perceptive fields of less than about 9~\AA{} struggle to distinguish between undoped and Al-doped structures. 
Either increasing the perceptive field or adding LES 
significantly enhances the accuracy of predicting this energy difference, closely matching DFT reference values of 934.8~meV for undoped and -66.9~meV for doped surfaces. Thus, although LES may not further improve the prediction for already highly accurate models, it assures that long-range effects associated with doping and adsorption are captured in all cases.

Notably, the CACE model with three-body interactions ($r=5.5$~\AA{} $n_l=1$ $\nu=2$) has the same body order and expressiveness as the HDNNPs~\cite{behler2007generalized}.
ko et al.~\cite{ko2021fourth} showed that the SR HDNNP fails to distinguish the energetics of the doped and undoped systems,
while the 4G-HDNNP that was trained on DFT partial charges and uses a charge equilibration scheme can reproduce the physical behavior.
Our observation is consistent with theirs, except that no training on DFT charges nor charge equilibration is needed.

\section{A long-range transferable MLIP for organics}

In the above, we benchmarked LES on specific systems and relatively small datasets.
Here, we explore whether the long-range correction scheme works at scale, for large and chemically diverse datasets.
This is not only relevant for validating the scaling capability of the method, but
is practically important for constructing emerging ``foundation models''~\cite{yuan2025foundation} that work across the periodic table.

The MACE-OFF23 (small (S), medium (M), and large (L))~\cite{kovacs2025MACEOFF} is a recent set of short-range transferable machine learning force fields for organic molecules, which are based on the MACE architecture and trained on finite organic systems:
small molecules of up to 50 atoms from about 85\% of the SPICE dataset version 1~\cite{eastman2023spice} with a neutral formal charge and containing ten chemical elements H, C, N, O, F, P, S, Cl, Br, and I,
as well as 50–90 atom molecules randomly selected from the QMugs dataset~\cite{isert2022qmugs}.
The dataset contains the energies and forces computed at the $\omega$B97M-D3(BJ)/def2-TZVPPD DFT level of accuracy.
We trained the MACELES-OFF model based on the same training/validation split of MACE-OFF23.

We note that our main purpose here is to showcase the effect of the LES augmentation compared to the baseline model, rather than fitting the most accurate model possible.
The MACELES-OFF model we trained used hyperparameters of cutoff radius $r=4.5$~\AA{}, chemical channels $k=192$, the order of the equivariant features $\ell=1$, and float32 precision.
The comparison of these hyperparameters with the original MACE-OFF23 models is presented in Table~\ref{tab:mace_models}.
Compared to the MACE-OFF23(S),
the MACELES-OFF has the same $r$ but higher $k$ and $\ell$.
Thanks to its lower precision as well as little overhead from LES,
the MACELES-OFF is about as fast as the MACE-OFF23(S),
and much faster than the MACE-OFF23(M) and MACE-OFF23(L) models, as shown in the timing benchmark in Fig.~\ref{fig:MACELES-OFF_timing} of Methods.

\begin{figure}
\centering
\includegraphics[width=\linewidth]{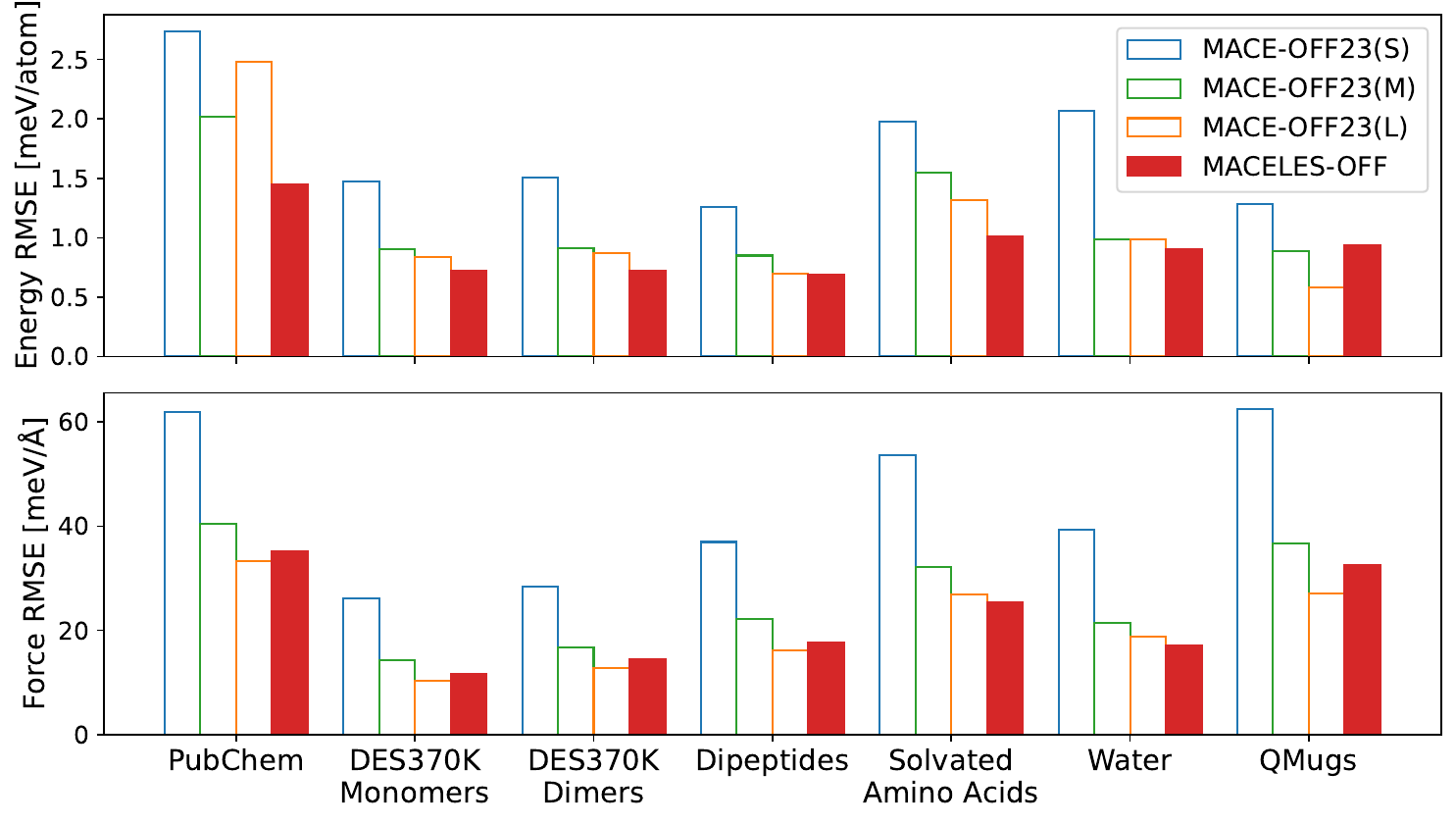}
     \caption{
    Comparison of test set root mean square errors (RMSEs) for energy and forces of the MACE-OFF23 (S, M, L) and MACELES-OFF models for organic molecules, with the underlying DFT reference values~\cite{kovacs2025MACEOFF}. Exact values are provided in Table~\ref{tab:macelesoff-error}.
    }
    \label{fig:MACELES-OFF_rmse}
\end{figure}

\begin{figure*}
\centering
\includegraphics[width=\linewidth]{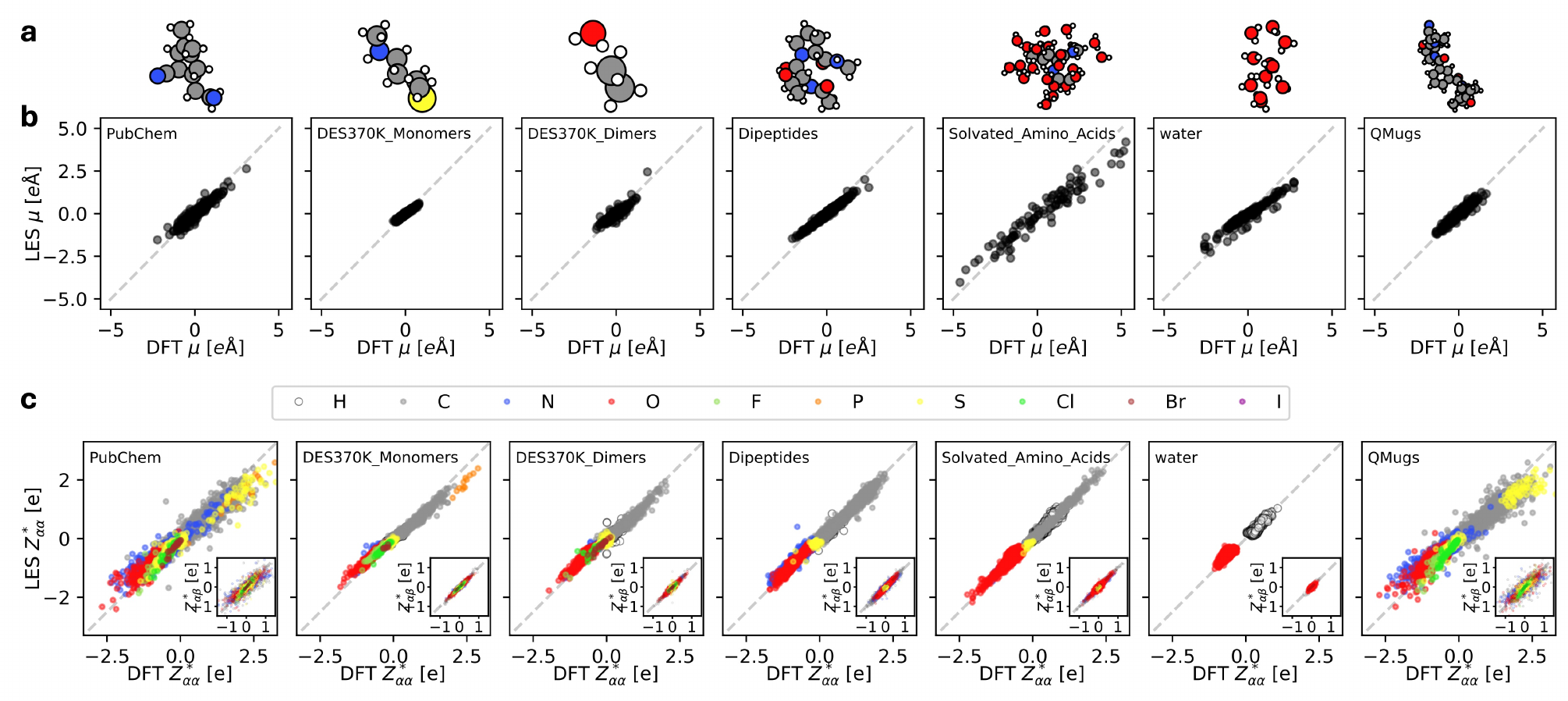}
     \caption{
     Assessment of dipole moments and Born effective charges (BECs) predicted by the MACELES-OFF for subsets of the MACE-OFF test set. 
     \figLabelCapt{a}: Representative configurations from each subset of the MACE-OFF test set. 
     \figLabelCapt{b}: Parity plots comparing predicted dipole moment components ($\mu$) against reference DFT values. 
     \figLabelCapt{c}: Parity plots comparing predicted BEC tensors against DFT reference. Data points are colored by atomic number, with diagonal BEC components ($Z^*_{\alpha\alpha}$) shown in the main plots and off-diagonal ($Z^*_{\alpha\beta}$) components shown in inset plots. 
    }
    \label{fig:macelesoff-bec}
\end{figure*}

In Fig.~\ref{fig:MACELES-OFF_rmse} we compare the RMSE values for energy and force for the three original MACE-OFF23 (S, M, L) and MACELES-OFF, on different subsets of test configurations (representative snapshots in Fig.~\ref{fig:macelesoff-bec}a).
Overall, the MACELES-OFF shows RMSEs comparable to the MACE-OFF23(L) that uses a larger cutoff $r=5$~\AA{} and higher $\ell=2$ (see Table~\ref{tab:mace_models}).
Compared to the MACE-OFF23(S) model that shares the same cutoff, the errors are about halved.
Amongst all the subsets, the MACELES-OFF shows particularly better accuracy for water clusters, solvated amino acids, and dipeptides, which may indicate that the model is well-suited for simulating biological systems.

To test whether LES is able to extract the correct electrostatics, we investigate if the model is able to infer the dipole moments $\mu$ (same as the polarization of finite systems) and BECs.
For about 50 configurations in each of the test subsets, we calculated reference BECs and dipole moments at the $\omega$B97m-D3(BJ)/Def2SVP level of theory (calculation details in Methods).
Fig.~\ref{fig:macelesoff-bec}b shows good agreement between the reference and the predicted dipole moments $\mu$.
Fig.~\ref{fig:macelesoff-bec}c compares the DFT and the BECs computed by LES via taking the derivative of the predicted $\mu$ as in Eqn.~\eqref{eq:z-finite},
showing broad agreement in both the on- and off-diagonal components over the broad range of elements and chemical species contained in the test set.
These agreements thus demonstrate that accurate inference of dipole moments and BECs is achievable in foundation-like models trained only on energy and forces.

\paragraph{Water}

A key requirement for organic force fields is the accurate description of bulk water.
Note that the training set only contains small, non-periodic water clusters; the application to the simulations of bulk liquid is in itself a generalization task.

\begin{figure}
\centering
\includegraphics[width=0.8\linewidth]{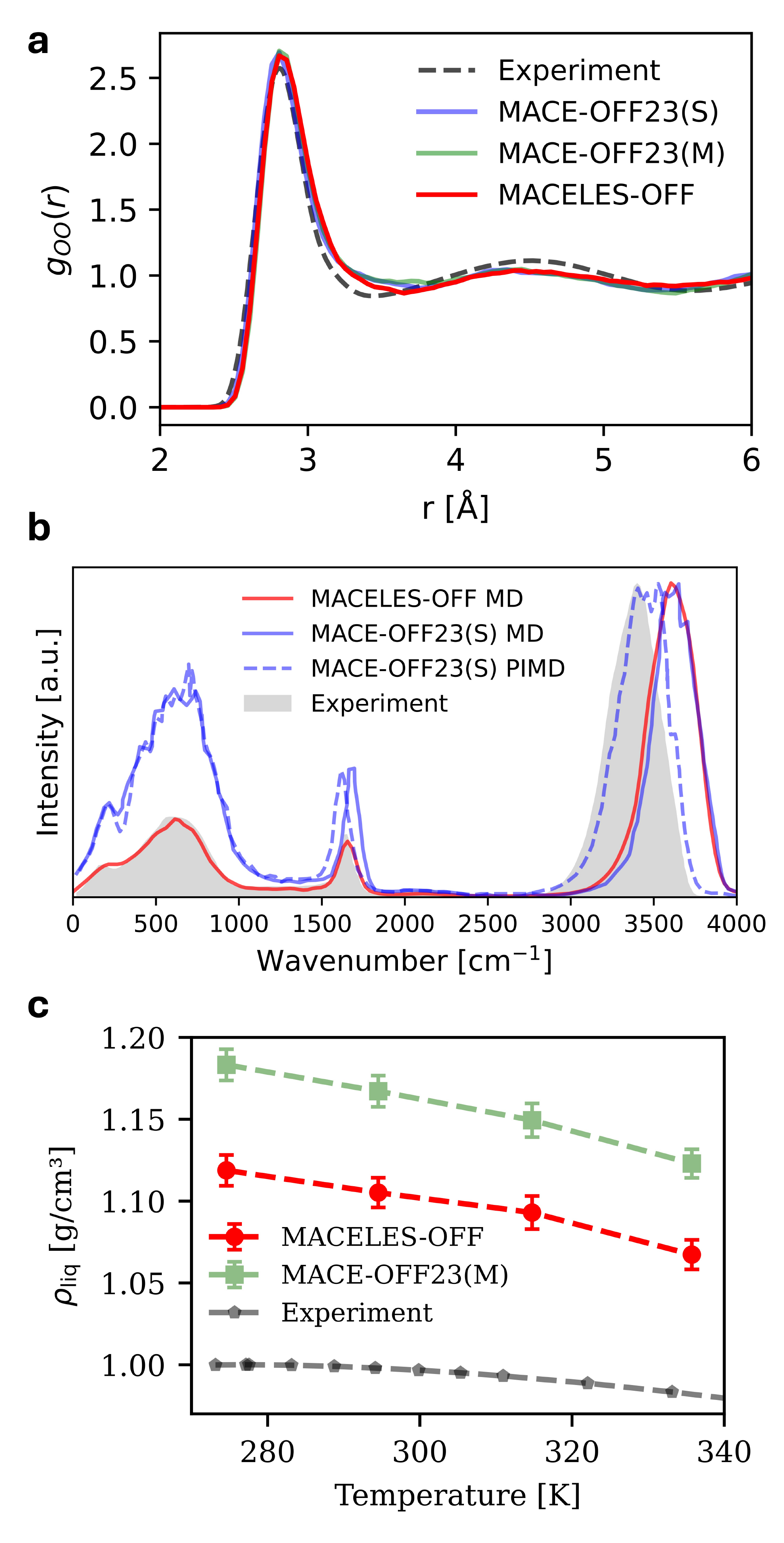}
    \caption{
    Bulk water properties predicted by the MACELES-OFF compared to the MACE-OFF23 and experimental data.
    \figLabelCapt{a}: Oxygen-Oxygen radial distribution functions in bulk water from the MACE-OFF23 models (S, M) and the MACELES-OFF model compared with experimental value (Ref.~\cite{skinner2014structure}).
    \figLabelCapt{b}: The infrared (IR) absorption spectra of bulk liquid water computed using the MACELES-OFF (red). For comparison, we also show the experimental IR spectrum~\cite{Bertie1996Infrared} (gray shading), the classical MD result from MACE-OFF23(S) (solid blue), and the result with quantum nuclear corrections to MACE-OFF23(S) (dashed blue), as reported in Ref.~\cite{kovacs2025MACEOFF}.
    \figLabelCapt{c}: Density isobar of liquid water at 1~atm. Both the MACELES-OFF (red circles) and the MACE-OFF23(M) (green squares) models show similar isobaric density characteristics for liquid water at 1~atm, with experimental data \cite{haynes2016crc} included for reference.
    }
    \label{fig:MACELES-OFF_water}
\end{figure}

We performed equilibrium NVT molecular dynamics simulations of bulk water at 300~K and experimental density, with further details provided in the Methods section.
Fig.~\ref{fig:MACELES-OFF_water}a demonstrates the capability of the MACELES-OFF model to accurately capture the radial distribution function (RDF) of water, and shows a result comparable to those from the previous MACE-OFF23 models (S, M) and experimental measurements~\cite{skinner2014structure}.

We computed an IR spectrum from the MD trajectory using predicted BECs, as
illustrated in Fig.~\ref{fig:MACELES-OFF_water}b, with additional details given in the Methods section. For comparison, we also included the IR spectrum predicted using MACE-OFF23(S), computed from the autocorrelation function of the time derivative of the total dipole moment predicted using a separate MACE-OFF23-$\mu$ model along classical MD trajectories~\cite{kovacs2025MACEOFF}, as well as experimental results~\cite{Bertie1996Infrared}.
The intramolecular bending mode ($\approx 1640~\mathrm{cm}^{-1}$), the intermolecular low-frequency libration band ($\approx 650~\mathrm{cm}^{-1}$), and the hydrogen-bond translational stretching mode ($\approx 200~\mathrm{cm}^{-1}$) are all well-captured by the MACELES-OFF,
whereas the MACE-OFF23(S) classical MD result reproduces the band shapes but not the intensities.
The predicted intramolecular vibrational mode (OH stretching band $\approx 3400~\mathrm{cm}^{-1}$) is blue-shifted relative to the experiment,
which is due to our use of classical MD without nuclear quantum effects (NQEs)~\cite{marsalek2017quantum}. To illustrate the impact of NQEs, we additionally show the IR spectrum obtained from the MACE-OFF23(S) with path-integral MD (PIMD) simulations within the PIGS approximation~\cite{musil2022Quantum}, as reported in Ref.~\cite{kovacs2025MACEOFF}. Incorporating NQEs brings the MACE-OFF23(S) frequencies into close agreement with experiment, especially by correcting the blue shift in the OH stretching region.

We simulated the temperature dependence of liquid water density based on NPT simulations conducted at 1~atm, as shown in Fig.~\ref{fig:MACELES-OFF_water}c. 
Comparative analysis demonstrates that MACELES-OFF achieves superior density prediction, exhibiting about 10\% overestimation of liquid water density compared to the original 18\% overestimation reported by the MACE-OFF23(M) model~\cite{kovacs2025MACEOFF}.
This improvement highlights the importance of long-range dipole-dipole interactions in predicting bulk properties such as density.
We also note that the MACE-OFF model can predict water density closer to experimental values by further increasing the cutoff to 6~\AA{}, which corresponds to a perceptive field of 12~\AA{}.

\begin{figure}
\centering
\includegraphics[width=0.8\linewidth]{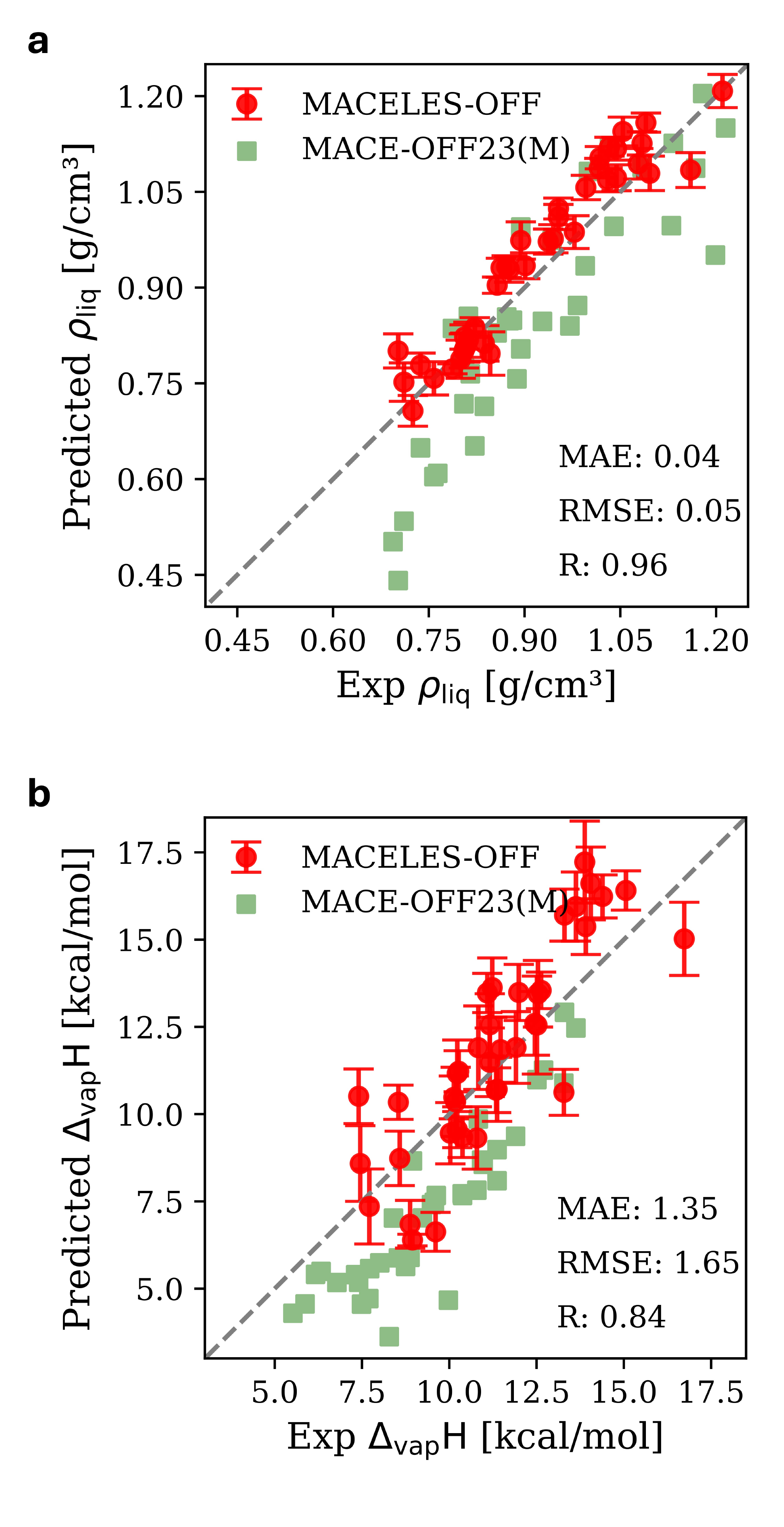}
    \caption{
    Predicted densities and heats of vaporization for diverse molecular liquids under ambient conditions (298~K and 1~atm) using the MACELES-OFF.
    \figLabelCapt{a}: Density ($\rho_{\rm liq}$) predictions from the MACELES-OFF (red circles) and the MACE-OFF23(M) (green squares), compared with experimental values~\cite{linstorm1998nist}.
    \figLabelCapt{b}: Heats of vaporization ($\Delta_{\mathrm{vap}} \mathrm{H}$) predictions from the MACELES-OFF (red circles) and the MACE-OFF23(M) (green squares), compared with experimental values~\cite{linstorm1998nist}.
    }
    \label{fig:mol_liquid}
\end{figure}

\paragraph{Molecular liquids}
We investigated the performance of MACELES-OFF in predicting condensed-phase properties such as densities ($\rho_{\rm liq}$) and enthalpy of vaporization ($\Delta_{\mathrm{vap}} \mathrm{H}$) under ambient conditions (298~K and 1~atm), as presented in \figref{fig:mol_liquid}. 
Reproducing such properties was observed to present challenges for short-ranged MLIPs~\cite{magduau2023machine, kovacs2025MACEOFF}.
We selected a comprehensive test set of 39 molecular liquids with boiling points greater than 320~K, referencing the MACEOFF benchmark~\cite{kovacs2025MACEOFF}.
This set encompasses diverse chemical classes highly relevant to chemistry and biology, including alcohols, amines, amides, ethers, aromatics/hydrocarbons, ketones/aldehydes, thiols/sulfides, and other functionalized compounds.

As shown in Fig.~\ref{fig:mol_liquid}, the MACELES-OFF model demonstrates superior accuracy in predicting liquid density and enthalpy of vaporization compared to the MACE-OFF23(M).
For liquid density, the MACELES-OFF achieves an MAE of 0.04~$\rm g/cm^3$ and RMSE of 0.05~$\rm g/cm^3$, which represents approximately half the error values of the MACE-OFF23(M) (MAE: 0.09~$\rm g/cm^3$, RMSE: 0.15~$\rm g/cm^3$). 
These low error values, coupled with a stronger correlation with experimental data ($R=0.97$) compared to the MACE-OFF23(M) model ($R = 0.89$), indicate that the MACELES-OFF accurately learned both short- and long-range intermolecular interactions.

Similarly, regarding the calculations of $\Delta_{\mathrm{vap}} \mathrm{H}$, the MACELES-OFF model yields a MAE of 1.32~kcal/mol and RMSE of 1.62~kcal/mol with a correlation of $R=0.84$ and thus outperforms the MACE-OFF23(M) model that has an MAE (2.18~kcal/mol) and RMSE (2.53~kcal/mol) with $R=0.87$. 
Moreover, while the MACE-OFF23(M) model exhibits a systematic $\Delta_{\mathrm{vap}} \mathrm{H}$ offset of about 2~kcal/mol, which is largely eliminated in the MACELES-OFF prediction. 

\paragraph{Simulations of biological molecules}

\begin{figure}[ht]
\centering
\includegraphics[width=\linewidth]{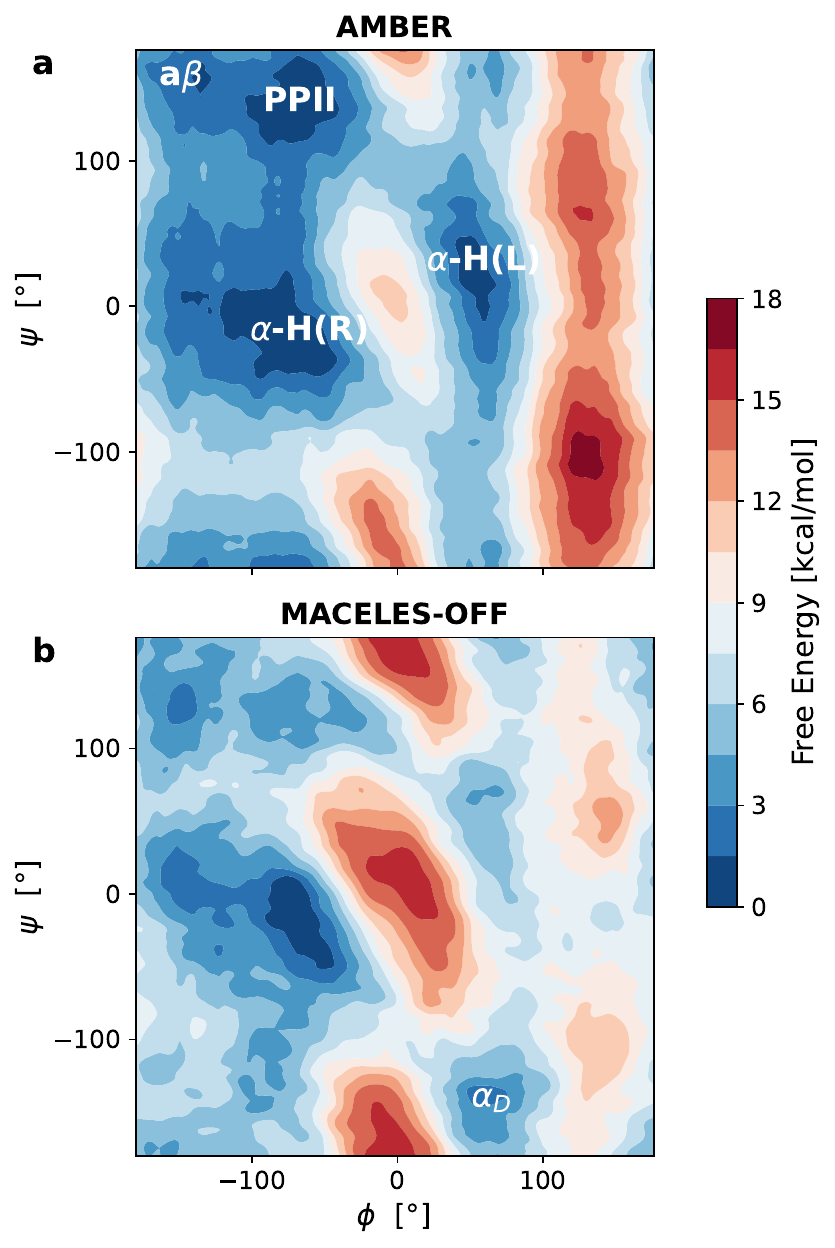}
    \caption{
    Torsional free energy surfaces (FES) of solvated alanine dipeptide computed at 310~K using different potentials.
    \figLabelCapt{a}: FES based on AMBER ff99SB/TIP3P.
    \figLabelCapt{b}: FES based on MACELES-OFF. 
    Four local minima are distinguished in the landscape and they correspond to distinct conformations defined by the central $\phi$ and $\psi$ backbone angles: antiparallel $\beta$-sheet (a$\beta$),  polyproline II (PPII),  right-handed $\alpha$-helix ($\alpha$-H(R)), and left-handed $\alpha$-helix structures ($\alpha$-H(L)). In addition, compared to AMBER ff99SB/TIP3P, MACELES-OFF identifies another local minimum that corresponds to the right-handed $\alpha$-helix region ($\alpha_D$). 
    }
    \label{fig:macelesoff-ala2}
\end{figure}

To probe the influence of the long-range interactions in MACELES-OFF on protein-protein and protein-water interactions, we studied two biological systems: solvated alanine dipeptide and the 1FSV protein~\cite{dahiyat_novo_1997}.

For the alanine dipeptide in water, we computed the free energy surface (FES) using well-tempered metadynamics~\cite{barducci_well-tempered_2008} simulations in explicit solvent (see Methods Section), as shown in Fig.~\ref{fig:macelesoff-ala2}. 
The AMBER ff99SB-ILDN force field exhibits an FES with four local minima, consistent with previous studies~\cite{zhang_molecular_2020, ple2025Foundation}: antiparallel $\beta$-sheet (a$\beta$), polyproline II (PPII), right-handed $\alpha$-helix ($\alpha$-H(R)), and left-handed $\alpha$-helix ($\alpha$-H(L)). In comparison, MACELES-OFF displays the deepest minimum at $\alpha$-H(R). The free energy difference between $\alpha$-H(R) and the second minimum (a$\beta$) is approximately 1.5~kcal/mol, while the difference with PPII is around 3.0~kcal/mol, consistent with the relative ordering of minima and similar to the relative free energies reported for MP2/cc-pVTZ level with implicit solvent~\cite{wang_solvation_2004}. 
Compared to the recent FENNIX-Bio1 model~\cite{ple2025Foundation} that includes long-range interactions, nuclear quantum effects, and is trained on the high-level SPICE2(+)-ccECP dataset, MACELES-OFF yields a very similar overall shape of the FES. 
Interestingly, the right-handed $\alpha$-helix region ($\alpha_D$ in Fig.~\ref{fig:macelesoff-ala2}b), 
%located at 50$^{\circ} <\phi < 100^{\circ}$ and $-180^{\circ} < \psi < -100^{\circ}$, 
does not appear as a clear minimum but rather as a shallow valley on the FES by both AMBER ff99SB-ILDN or AMBER99~\cite{ple2025Foundation}, likely due to insufficient treatment of water-protein interactions, an issue well tackled by MACELES-OFF.

\begin{figure}
\centering
\includegraphics[width=\linewidth]{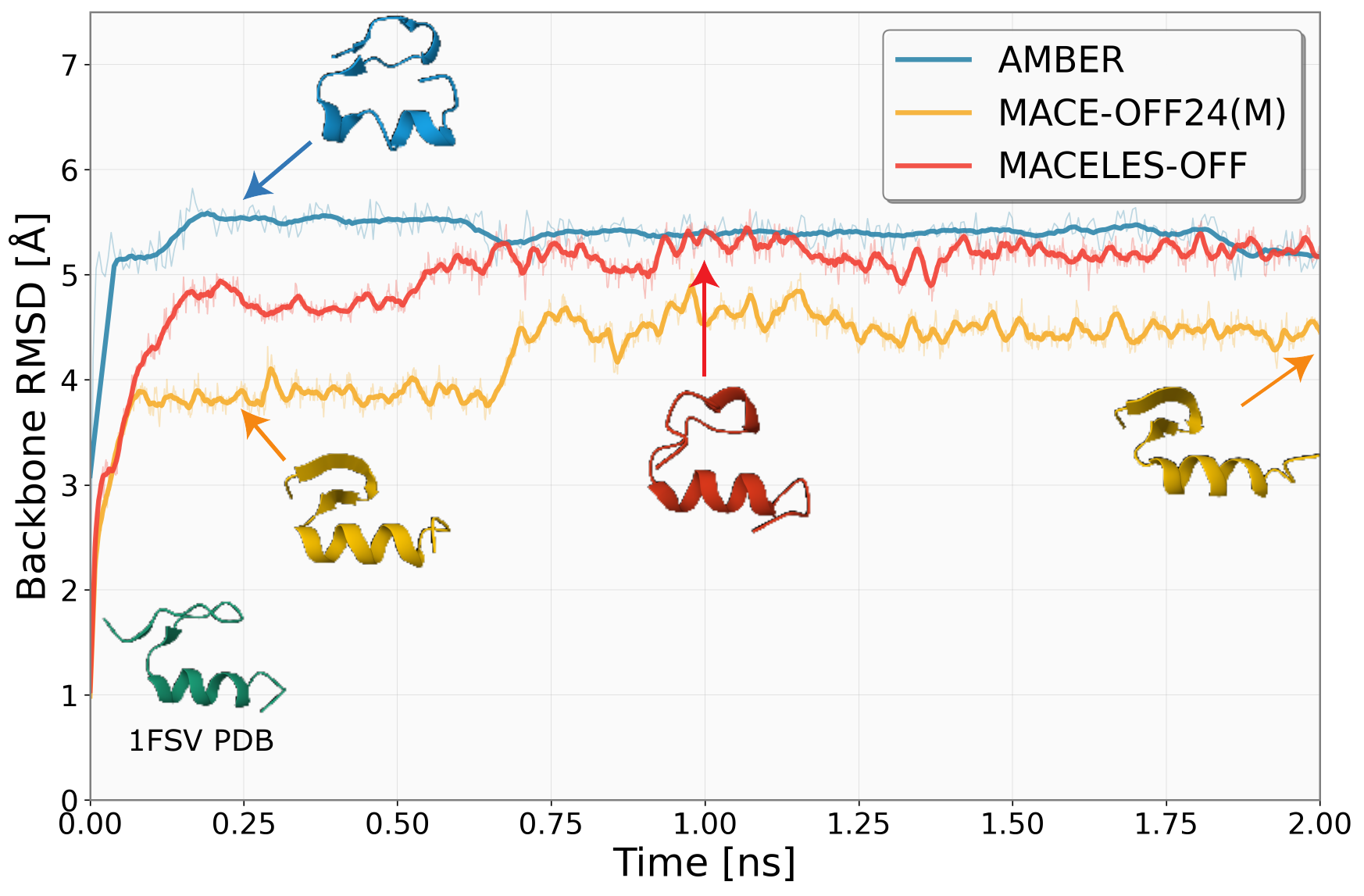}
     \caption{
      The root mean square deviation (RMSD) of the 1FSV protein during a 2~ns gas-phase NVT MD simulation at 310~K. Time evolution of RMSD and protein structures compared to the initial PDB structure for AMBER ff99SB (blue), MACEOFF-24(M) (orange), and MACELES-OFF (red).
      }
    \label{fig:macelesoff-1fsv}
\end{figure}

For the 1FSV protein, we performed 2~ns MD simulations in vacuum for all three models, starting from the crystallographic PDB structure. The 1FSV protein poses a significant challenge due to its 10 positively and 5 negatively charged groups. As shown in Fig.~\ref{fig:macelesoff-1fsv}, both MACELES-OFF and MACE-OFF24(M) began folding the protein into a more compact conformation, characteristic of gas-phase behavior. However, their trajectories diverged over time. MACELES-OFF exhibited more stable dynamics, remaining close to AMBER ff99SB-ILDN~\cite{lindorff-larsen_improved_2010}, with a root-mean-square deviation (RMSD) of 5.14~\AA{} with respect to the 1FSV PDB structure, just 0.085~\AA{} lower than AMBER’s 5.22~\AA{}. In contrast, MACE-OFF24(M) showed continued RMSD drift and over-compaction, reaching 4.43~\AA{}.

Further structural analysis confirmed these trends through the calculation of salt bridges and hydrogen bond networks. MACE-OFF24(M) exhibited the highest number of salt bridges (159, +35\% vs. PDB) and hydrogen bonds (286, +55\%), indicating an over-collapsed structure. By comparison, MACELES-OFF formed 149 salt bridges and 263 hydrogen bonds, values more consistent with AMBER (143 and 258, respectively).

Overall, the results above suggest that MACELES-OFF can provide a good description of the FES of a small peptide solvated in water.
Comparing MACELES-OFF and MACEOFF for the folding of 1FSV in vacuum, long-range electrostatics play a critical role in governing the dynamics and structural stability of the protein.
These findings underscore the importance of including long-range interactions in simulations of biomolecular systems.

\section{Discussion}

We present LES as a ``universal'' augmentation framework for adding long-range electrostatics to short-range MLIPs. 
It is universal in three ways:
First, LES can work together with many short-ranged MLIPs in an architecture-agnostic way as long as there are local invariant features for atoms.
We patched LES to MACE, NequIP, CACE, and MatGL, and it is our ongoing commitment to include LES in other MLIPs, making it more widely accessible.
Interestingly, LES works well with the CACE models with three-body interactions $\nu=2$, which are mathematically equivalent to the earlier generation of MLIPs (e.g., HDNNPs~\cite{behler2007generalized}, Gaussian Approximation Potentials~\cite{bartok2010gaussian}, and DeepMD~\cite{zhang2022deep}).
This suggests that LES will also complement well those lower-body-order MLIP methods.
The LES-augmented HDNNPs would indeed be quite similar to
the third-generation HDNNPs (3G-HDNNPs)~\cite{ko2021fourth}, except that not explicitly trained on partial DFT charges.
The fact that 3G-HDNNPs have poor performance in many systems~\cite{ko2021fourth} may be attributed to the explicit charge training, rather than inferring from energy and forces.

Second, the LES augmentation is useful 
across a wide range of systems.
Across the three benchmarks (bulk water, dipeptides, and
Au$_{2}$ on MgO(001)), the inclusion of LES yields consistent, architecture-agnostic improvements.
Even when the reduction of energy and force RMSEs is modest, LES significantly improves physical observables.
For example, in the water and dipeptide set, LES achieves accurate BEC predictions despite training exclusively on energy and forces.
BECs are not only well-defined physical quantities that serve as sensitive probes of electrostatics, but are also critical in calculating a range of electrical response properties such as dielectric constants, ionic conductivities, and dipole correlation functions~\cite{zhong2025machine}. 

Third, LES can be scaled to large datasets with chemical diversity and can be used to train universal MLIPs.
As a demonstration, the MACELES-OFF model shows better accuracy compared to the short-ranged baseline models,
encodes the correct electrostatics,
and exhibits better transferability.
For instance, the MACE-OFF23 models show an amazingly wide range of accurate predictions across small molecules, biological systems, and molecular crystals,
but they are less accurate in predicting the densities and heats of vaporization of molecular liquids.
Such shortcomings are attributed to the lack of long-range interactions~\cite{kovacs2025MACEOFF}.
Indeed, incorporating LES effectively solved such issues and provides much improved predictions for the liquids.
In addition, electrical response properties such as IR spectra are readily available.

Additionally, using LES only requires standard energy and force labels. This is an advantage for scaling up the training of universal MLIPs with large-scale datasets,
as most such datasets~\cite{Jain2013, levine2025openmolecules2025omol25, tran2023Open, barroso_omat24}, especially of extended systems, do not contain explicit polarization or charge information.
However, the dipole moments of gas-phase molecules are more readily available,
and in principle one can include these in the training of partial charges under the LES framework,
by adding a dipole loss function as in 
AIMNET2~\cite{Anstine2025}, SO3LR~\cite{Kabylda2025}, and MPNICE~\cite{weber2025efficient}.
Towards true foundation models with full electrostatic physics,
one also needs to address the issue of the field-dependence of atomic charges, which LES currently does not handle, but we have ongoing work along this direction.

To conclude, by incorporating long-range electrostatics and without training on labels other than energy, forces (and stresses), the LES method addresses a core limitation of current MLIPs.
The LES library works as a drop-in patch for many short-ranged MLIPs,
and demonstrates reliable and consistent performance across a wide range of architectures, chemical system types, and data sizes.

\section{Methods}

\subsection{Implementation}
LES is implemented in PyTorch~\cite{Paszke2019PyTorch} and is easily compatible with MLIP packages also implemented in PyTorch.
In CACE~\cite{cheng2024cartesian}, we implemented a \texttt{LESwrapper}, and we note that CACE has its own native LES implementation~\cite{Cheng2025Latent} which yields consistent results.
In MACE~\cite{batatia2022mace}, we added a \texttt{MACELES} model which can be used in the same way as the original \texttt{MACE} model, and all training and evaluation procedures stay the same.
In NequIP~\cite{batzner20223}, we extended the \texttt{FullNequIPGNNEnergyModel} by adding a  \texttt{use\_les} key to optionally activate the \texttt{LatentEwaldSum} and the \texttt{AddEnergy} modules; all training and evaluation procedures remain unchanged. To compute the BECs after training, one can add the \texttt{ToggleLESCallback} to the \texttt{callbacks} list and enable the \texttt{compute\_bec} setting.  
In MatGL~\cite{ko2025_matgl}, the inference and training are treated in different modules. We extended the \texttt{Potential} module with \texttt{calc\_BEC = True} to include LES and autograd to compute the BECs for the inference. We extended the \texttt{PotentialLightningModule} with \texttt{include\_long\_range = True} to call \texttt{Potential} with long-range interactions for the training.

\subsection{Details for benchmarks}

The training scripts with hyperparameters as well as the trained models for the water, dipeptides, and Au$_{2}$ on MgO(001) datasets are included in the SI repository.
For the LES augmentation part, we always use the default parameters: 
$\sigma = 1$~\AA{}, $dl=2$~\AA{} (which corresponds to a $\mathbf{k}$-point cutoff of $k_c = \pi$ in the Ewald summation).
%\bc{anything worth mentioning here? Any tricks on the training or things to note?}

\begin{figure}[t]
\centering
\includegraphics[width=0.95\linewidth]{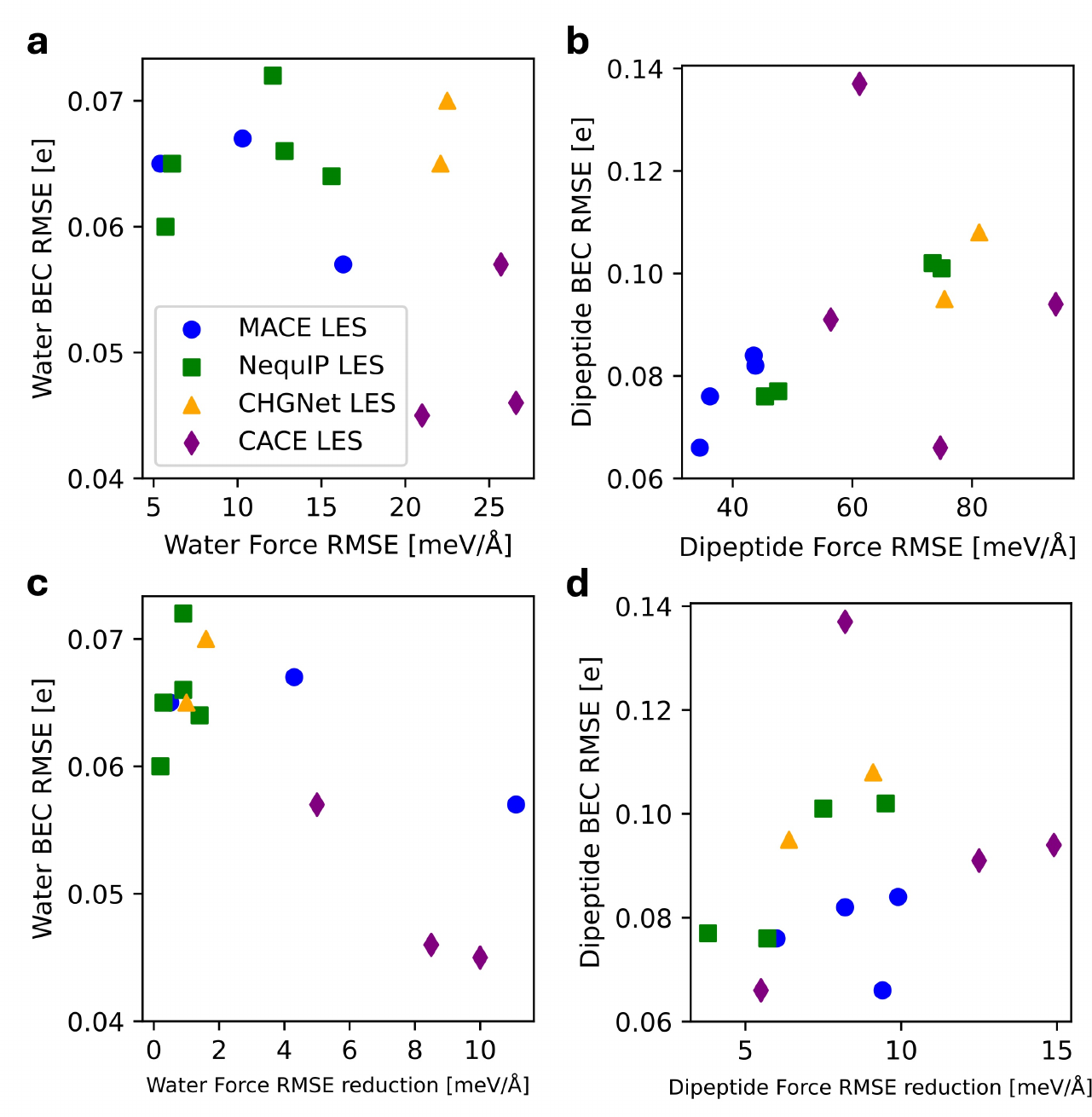}
     \caption{
     Correlation analysis between force and Born effective charge (BEC) RMSEs for LES-augmented MLIPs.
     \figLabelCapt{a, b}: Correlation between force RMSE and BEC RMSE for water (a) and dipeptide (b) systems.
     \figLabelCapt{c, d}: Correlation between force RMSE reduction after LES augmentation and BEC RMSE for water (c) and dipeptide (d) systems.
    }
    \label{fig:method-force_bec_correlation}
\end{figure}

Fig.~\ref{fig:method-force_bec_correlation} shows the relation between force RMSE and BEC RMSE for LES-augmented MLIPs (panels a and b),
and the relation between the reduction in force RMSE after LES augmentation and the corresponding BEC RMSE (panels c and d), for water and dipeptide systems. 

\begin{table}[t]
\centering
\begin{tabular}{|l|l|r|r|}
\hline
\textbf{MLIP} & \textbf{Hyperparameters} & \textbf{Undoped} & \textbf{Doped} \\
\hline
MACE & $r=5.5$~\AA{}, $n_{l}=1$        & 419.0 & 419.0 \\
MACE LES & $r=5.5$~\AA{}, $n_{l}=1$  & 919.3 & -56.5 \\
NequIP & $r=4.0$~\AA{}, $n_{l}=2$, $\ell$=1  & 390.2 & 390.3 \\
NequIP LES & $r=4.0$~\AA{} $n_{l}=2$, $\ell$=1  & 973.3 & -45.1 \\
CHGNet & $r=4.5$~\AA{}, $n_{l}=1$  & 411.9 & 411.9 \\
CHGNet LES & $r=4.5$~\AA{}, $n_{l}=1$  & 925.5 & -103.5 \\
CACE & $r=5.5$~\AA{}, $n_{l}=1$, $\nu$=2 & 421.5 & 421.4 \\
CACE LES& $r=5.5$~\AA{}, $n_{l}=1$, $\nu$=2 & 931.7 & -72.2 \\
CACE & $r=5.5$~\AA{}, $n_{l}=1$, $\nu$=3 & 430.9 & 430.9 \\
CACE LES& $r=5.5$~\AA{}, $n_{l}=1$, $\nu$=3 & 930.9  & -71.1 \\

\hline
\multicolumn{2}{|c|}{\textbf{DFT reference}}            & 934.8 & -66.9 \\
\hline
\end{tabular}
\caption{Energy difference ($E_\mathrm{wetting}-E_\mathrm{non-wetting}$) in meV between the wetting and non-wetting configurations for doped and undoped substrates.}
\label{tab:wetting-energy}
\end{table}

Table~\ref{tab:wetting-energy} summarizes the energy differences (see Fig.~\ref{fig:benchmark-Au_MgO}) between wetting and non-wetting configurations for doped and undoped substrates in the Au$_{2}$-MgO(001) system.

\subsection{Details on training the MACELES-OFF model}

The training scripts as well as the trained models for the MACELES-OFF are included in the SI repository.
The hyperparameters are summarized and compared with the original MACE-OFF models in Table~\ref{tab:mace_models}.

\begin{table}[ht]
    \centering
    \resizebox{\linewidth}{!}{%
    \begin{tabular}{l|c|c|c|c|c|c}
        \toprule \hline\hline
        & 23(S) & 23(M) & 23(M)b & 23(L) & 24(M) & LES  \\ \hline
        \toprule
        Cutoff $r$ (\AA{}) & 4.5 & 5.0 & 6.0 & 5.0 & 6.0 & 4.5\\ \hline
        Channels $k$ & 96 & 128 & 128 & 192 & 128 & 192 \\ \hline
        Irrep order $\ell$  & 0 & 1 & 1 & 2 & 1 & 1 \\ \hline
        SPICE version & 1 & 1 & 1 & 1 & 2 & 1 \\ \hline
        Precision & float64 & float64 & float64 & float64 & float64 & float32 \\ \hline\hline
        \bottomrule
    \end{tabular}
    }
    \caption{
    Hyperparameters of the MACE-OFF and MACELES-OFF models. All MACE-OFF models (23(S), 23(M), 23(L), and 24(M)) are taken from Ref.~\cite{kovacs2025MACEOFF}. 24(M) was trained on the updated SPICE version 2~\cite{eastman2024SPICE}. 
    }
    \label{tab:mace_models}
\end{table}

\begin{table*}

\definecolor{myblue}{RGB}{170,200,255}
\definecolor{mygreen}{RGB}{170,255,200}
\definecolor{myyellow}{RGB}{255,255,170}
\definecolor{myred}{RGB}{255,170,200}
\centering
\small

\begin{tabular}{c|c||cccccccc}
\hhline{~~~-------}
\hhline{~~~-------}
\hhline{~~~-------}
\multicolumn{3}{c}{} & PubChem & \makecell{DES370K\\Monomers} & \makecell{DES370K\\Dimers} & Dipeptides & \makecell{Solvated\\Amino Acids} & Water & QMugs \\
% & Tripeptide \\
\hline
\hline
\multirow{9}{*}{\rotatebox[origin=c]{90}{MAE}} & \multirow{3}{*}{\makecell{$\rm E$\\$(\rm meV/at)$}} & 23S & \cellcolor{myblue!90} \textcolor{black} {1.41} & \cellcolor{myblue!90} \textcolor{black} {1.04} & \cellcolor{myblue!90} \textcolor{black} {0.98} & \cellcolor{myblue!90} \textcolor{black} {0.84} & \cellcolor{myblue!90} \textcolor{black} {1.60} & \cellcolor{myblue!90} \textcolor{black} {1.67} & \cellcolor{myblue!90} \textcolor{black} {1.03} \\
 & & 23M & \cellcolor{mygreen!100} \textcolor{black} {0.91} & \cellcolor{mygreen!100} \textcolor{black} {0.63} & \cellcolor{mygreen!100} \textcolor{black} {0.58} & \cellcolor{mygreen!100} \textcolor{black} {0.52} & \cellcolor{mygreen!100} \textcolor{black} {1.21} & \cellcolor{mygreen!100} \textcolor{black} {0.76} & \cellcolor{mygreen!100} \textcolor{black} {0.69} \\
 & & 23L & \cellcolor{myyellow!90} \textcolor{black} {0.88} & \cellcolor{myyellow!90} \textcolor{black} {0.59} & \cellcolor{myyellow!90} \textcolor{black} {0.54} & \cellcolor{myyellow!90} \textcolor{black} {0.42} & \cellcolor{myyellow!90} \textcolor{black} {0.98} & \cellcolor{myyellow!90} \textcolor{black} {0.83} & \cellcolor{myyellow!90} \textcolor{black} {0.45} \\

  & & \textcolor{red}{LES} & \cellcolor{myred!90} \textcolor{black} {0.71} & \cellcolor{myred!90} \textcolor{black} {0.54} & \cellcolor{myred!90} \textcolor{black} {0.47} & \cellcolor{myred!90} \textcolor{black} {0.52} & \cellcolor{myred!90} \textcolor{black} {0.82} & \cellcolor{myred!90} \textcolor{black} {0.69} & \cellcolor{myred!90} \textcolor{black} {0.76} \\
  
\hhline{~---------}
\hhline{~---------}
 & \multirow{3}{*}{\makecell{$\rm F$\\$(\rm meV/\AA)$}} & 23S & \cellcolor{myblue!90} \textcolor{black} {35.68} & \cellcolor{myblue!90} \textcolor{black} {17.63} & \cellcolor{myblue!90} \textcolor{black} {16.31} & \cellcolor{myblue!90} \textcolor{black} {25.07} & \cellcolor{myblue!90} \textcolor{black} {38.56} & \cellcolor{myblue!90} \textcolor{black} {28.53} & \cellcolor{myblue!90} \textcolor{black} {41.45} \\
 & & 23M & \cellcolor{mygreen!100} \textcolor{black} {20.57} & \cellcolor{mygreen!100} \textcolor{black} {9.36} & \cellcolor{mygreen!100} \textcolor{black} {9.02} & \cellcolor{mygreen!100} \textcolor{black} {14.27} & \cellcolor{mygreen!100} \textcolor{black} {23.26} & \cellcolor{mygreen!100} \textcolor{black} {15.27} & \cellcolor{mygreen!100} \textcolor{black} {23.58} \\
 & & 23L & \cellcolor{myyellow!90} \textcolor{black} {14.75} & \cellcolor{myyellow!90} \textcolor{black} {6.58} & \cellcolor{myyellow!90} \textcolor{black} {6.62} & \cellcolor{myyellow!90} \textcolor{black} {10.19} & \cellcolor{myyellow!90} \textcolor{black} {19.43} & \cellcolor{myyellow!90} \textcolor{black} {13.57} & \cellcolor{myyellow!90} \textcolor{black} {16.93} \\

  & & \textcolor{red}{LES} & \cellcolor{myred!90} \textcolor{black} {17.24} & \cellcolor{myred!90} \textcolor{black} {7.65} & \cellcolor{myred!90} \textcolor{black} {7.48} & \cellcolor{myred!90} \textcolor{black} {11.46} & \cellcolor{myred!90} \textcolor{black} {18.37} & \cellcolor{myred!90} \textcolor{black} {12.54} & \cellcolor{myred!90} \textcolor{black} {20.11} \\
  
\hline
\hline

\multirow{9}{*}{\rotatebox[origin=c]{90}{RMSE}} & \multirow{3}{*}{\makecell{$\rm E$\\$(\rm meV/at)$}} & 23S & \cellcolor{myblue!90} \textcolor{black} {2.74} & \cellcolor{myblue!90} \textcolor{black} {1.47} & \cellcolor{myblue!90} \textcolor{black} {1.51} & \cellcolor{myblue!90} \textcolor{black} {1.26} & \cellcolor{myblue!90} \textcolor{black} {1.98} & \cellcolor{myblue!90} \textcolor{black} {2.07} & \cellcolor{myblue!90} \textcolor{black} {1.28} \\
 & & 23M & \cellcolor{mygreen!100} \textcolor{black} {2.02} & \cellcolor{mygreen!100} \textcolor{black} {0.90} & \cellcolor{mygreen!100} \textcolor{black} {0.91} & \cellcolor{mygreen!100} \textcolor{black} {0.85} & \cellcolor{mygreen!100} \textcolor{black} {1.55} & \cellcolor{mygreen!100} \textcolor{black} {0.99} & \cellcolor{mygreen!100} \textcolor{black} {0.89} \\
 % \textcolor{black} {0.75} \\
 & & 23L & \cellcolor{myyellow!90} \textcolor{black} {2.48} & \cellcolor{myyellow!90} \textcolor{black} {0.84} & \cellcolor{myyellow!90} \textcolor{black} {0.87} & \cellcolor{myyellow!90} \textcolor{black} {0.70} & \cellcolor{myyellow!90} \textcolor{black} {1.32} & \cellcolor{myyellow!90} \textcolor{black} {0.99} & \cellcolor{myyellow!90} \textcolor{black} {0.58} \\

   & & \textcolor{red}{LES} & \cellcolor{myred!90} \textcolor{black} {1.45} & \cellcolor{myred!90} \textcolor{black} {0.72} & \cellcolor{myred!90} \textcolor{black} {0.72} & \cellcolor{myred!90} \textcolor{black} {0.69} & \cellcolor{myred!90} \textcolor{black} {1.01} & \cellcolor{myred!90} \textcolor{black} {0.90} & \cellcolor{myred!90} \textcolor{black} {0.94} \\
   
\hhline{~---------}
\hhline{~---------}
 & \multirow{3}{*}{\makecell{$\rm F$\\$(\rm meV/\AA)$}} & 23S & \cellcolor{myblue!90} \textcolor{black} {61.83} & \cellcolor{myblue!90} \textcolor{black} {26.15} & \cellcolor{myblue!90} \textcolor{black} {28.48} & \cellcolor{myblue!90} \textcolor{black} {36.97} & \cellcolor{myblue!90} \textcolor{black} {53.55} & \cellcolor{myblue!90} \textcolor{black} {39.33} & \cellcolor{myblue!90} \textcolor{black} {62.46} \\
 & & 23M & \cellcolor{mygreen!100} \textcolor{black} {40.51} & \cellcolor{mygreen!100} \textcolor{black} {14.31} & \cellcolor{mygreen!100} \textcolor{black} {16.81} & \cellcolor{mygreen!100} \textcolor{black} {22.25} & \cellcolor{mygreen!100} \textcolor{black} {32.19} & \cellcolor{mygreen!100} \textcolor{black} {21.40} & \cellcolor{mygreen!100} \textcolor{black} {36.73} \\
 % {33.94} \\
 & & 23L & \cellcolor{myyellow!90} \textcolor{black} {33.34} & \cellcolor{myyellow!90} \textcolor{black} {10.27} & \cellcolor{myyellow!90} \textcolor{black} {12.81} & \cellcolor{myyellow!90} \textcolor{black} {16.19} & \cellcolor{myyellow!90} \textcolor{black} {26.91} & \cellcolor{myyellow!90} \textcolor{black} {18.78} & \cellcolor{myyellow!90} \textcolor{black} {27.17} \\

   & & \textcolor{red}{LES} & \cellcolor{myred!90} \textcolor{black} {35.26} & \cellcolor{myred!90} \textcolor{black} {11.70} & \cellcolor{myred!90} \textcolor{black} {14.42} & \cellcolor{myred!90} \textcolor{black} {17.62} & \cellcolor{myred!90} \textcolor{black} {25.31} & \cellcolor{myred!90} \textcolor{black} {17.08} & \cellcolor{myred!90} \textcolor{black} {32.57} \\
   
\hline
\hline
\end{tabular}
\caption{
Test set root mean square errors (RMSEs) for energy and forces of the MACE-OFF23 (S, M, L) and MACELES-OFF models for organic molecules compared to the underlying DFT reference data~\cite{kovacs2025MACEOFF}.
}
\label{tab:macelesoff-error}
\end{table*}

Table~\ref{tab:macelesoff-error} summarizes the test errors for each dataset. For the PubChem dataset, 13 outlier configurations were removed from a total of 34,093 test structures in accordance with the MACE-OFF paper~\cite{kovacs2025MACEOFF}.

Reference Born effective charges (BECs) and dipole moments for the MACELES-OFF were computed in PySCF~\cite{sun2018pyscf} using the $\omega$B97m-D3(BJ) functional in the Def2SVP basis set. Code for calculating BECs can be found in the infrared module of the properties extension of PySCF \url{https://github.com/pyscf/properties.git}.

\subsection{Details on MD simulations using the MACELES-OFF model}

\paragraph{Water} 
We performed equilibrium NVT simulations of bulk water using the MACE-OFF23 (S, M) and MACELES-OFF models in ASE \cite{hjorth_larsen_atomic_2017} at a density of 0.997~g/cm$^{3}$ and 300~K for a system of 64 water molecules, employing the Nos\'e-Hoover thermostat for Fig.~\ref{fig:MACELES-OFF_water}a and b. Each MD simulation was run for 200,000 steps with a timestep of 0.25~fs, corresponding to 50~ps.

The time-dependent current of polarization of the system is given by 
$\mathbf{J}(t) = \sum_{i=1}^N Z_i^*(t)\cdot\mathbf{v}_i(t)$, where $Z_i^*(t)$ is the BEC tensor and $\mathbf{v}_i(t)$ is the velocity of an atom $i$.
The IR spectra are obtained via the Fourier transform of the current-current autocorrelation function,
\begin{equation}
    I(\omega) \propto \int_0^T dt \left\langle \mathbf{J}(0) \mathbf{J}(t) \right\rangle e^{-i\omega t}.
    \label{eq:ir}
\end{equation}
The IR spectrum for Fig.~\ref{fig:MACELES-OFF_water}b was calculated from the MACELES-OFF MD trajectory, and the resulting raw intensity was smoothed by convolving the spectrum with a Gaussian kernel.

We simulated the density isobar of liquid water at 1~atm by investigating four temperatures: 275~K, 295~K, 315~K, and 336~K. These NPT simulations employed a system of 64 water molecules, with a 1.0~fs timestep over 300~ps of simulation time, utilizing the Nos\'e-Hoover thermostat and Berendsen barostat implemented in ASE.

\paragraph{Molecular liquids} 
For the preparation of MD simulations in the liquid phase, SMILES strings for the organic molecules were first obtained from the PubChem database~\cite{kim2025pubchem} and subsequently converted to PDB format using the Python package RDKit~\cite{landrum2025rdkit}. 
The initial molecular configurations were generated at 60\% of experimental density using the GROMACS command-line interface~\cite{Berendsen1995GROMACS}, followed by structural pre-optimization with the General AMBER Force Field (GAFF)~\cite{wang2006automatic}.
A two-stage optimization protocol was then implemented: geometry optimization using the MACELES-OFF model with the BFGS algorithm in ASE, followed by NVT simulations for 5,000~steps using a Nos\'{e}-Hoover thermostat with a timestep of 1.0~fs to achieve structural relaxation.
Production NPT simulations were performed for 300~ps using a 1.0~fs timestep, using the same Nos\'{e}-Hoover thermostat coupled with the Martyna-Tobias-Klein (MTK) barostat \cite{Yoshida1990MTK}. 
Final density and potential energy values were computed by averaging over the last 150 ps to ensure proper equilibration. 
To compute the enthalpy of vaporization, potential energies for the gas phase were obtained based on simulations of isolated molecules in a non-periodic box.
The ideal gas approximation was employed to account for pressure effects in the gas phase.

It should also be noted the 1,1,2,2-tetrachloroethane system vaporized under NPT simulation using MACELES-OFF, although it remained liquid using MACE-OFF23(M).
This might be due to the difficulty of preparing the initial condition of this system.
Despite this exception, the MACELES-OFF demonstrated successful simulation of another chlorinated hydrocarbon, 1,4-dichlorobutane, a structurally analogous compound that was an outlier in the MACE-OFF23(M) predictions of $\Delta_{\mathrm{vap}} \mathrm{H}$.
Furthermore, the MACELES-OFF accurately predicted both the density and enthalpy of vaporization for chlorinated aniline, thereby further proving its capacity to model intra- and intermolecular interactions involving chlorine atoms.
 
Simulated density, enthalpy of vaporization, and optimized initial configurations for both liquid and gas phases are provided in the SI repository.

\paragraph{Simulations of biological molecules} 
The PDB structure of alanine dipeptide was solvated in a cubic box of 28~\AA{} using the \texttt{addSolvent} method in OpenMM, resulting in a system of 2,050 atoms. The system was modeled using the AMBER ff99SB-ILDN force field~\cite{lindorff-larsen_improved_2010} and the TIP3P water model~\cite{jorgensen_comparison_1983}. The structure was first geometry-optimized with a tolerance of 10~kJ/mol/nm, followed by 50~ps of NPT equilibration at 310~K and 1.013~bar using a Monte Carlo barostat and Langevin thermostat, with a timestep of 1.0~fs and a friction coefficient of 1.0~ps$^{-1}$. Production NVT simulations with well-tempered metadynamics were performed for 2~ns for MACELES-OFF and 5~ns for AMBER ff99SB-ILDN/TIP3P along the central $\phi$ and $\psi$ backbone angles using ASE patched with the PLUMED code (version 2.9)~\cite{bonomi_promoting_2019, tribello_plumed_2014}. 

For both collective variables ($\phi$ and $\psi$), we employed a Gaussian height of 2.5~kJ/mol and a bias factor of 30. Gaussian potentials were deposited on the energy landscape every 500 steps, and their width was adapted on the fly based on the local diffusivity \cite{branduardi2012metadynamics}. The FES was computed from reweighting.

For the 1FSV protein, the PDB structure~\cite{dahiyat_novo_1997} was placed in a cubic simulation box at a density of 0.6~g/cm$^3$ to avoid interatomic overlaps and to maintain periodic boundary conditions as required by the LES framework. Geometry optimization was performed in three successive stages using the FIRE optimizer~\cite{Bitzek2006_FIRE}, with gradually decreasing convergence criteria down to a maximum force of 0.2~eV/\AA{}. We then ran 2~ns of MD production in the NVT ensemble at 310~K using a 1~fs timestep and a friction coefficient of 0.1~ps$^{-1}$, starting from the PDB crystallographic structure.

Salt bridge analysis was performed by counting the number of interactions within a 4.0~\AA{} cutoff between the side-chain oxygen atoms of acidic residues (e.g., ASP) and the nitrogen atoms of basic residues (e.g., ARG). Similarly, hydrogen bonds were assessed by counting interactions within a 3.5~\AA{} cutoff between donor and acceptor atoms.

\paragraph{Timing} 

\begin{figure}
\centering
\includegraphics[width=\linewidth]{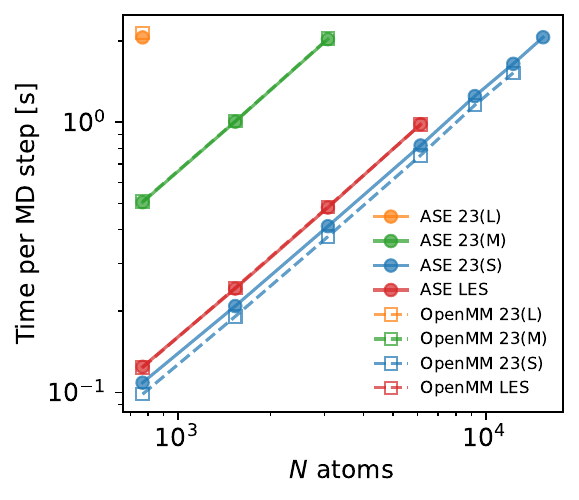}
     \caption{
     Computational performance benchmarks of molecular dynamics (MD) simulations using the MACELES-OFF and the MACE-OFF23~\cite{kovacs2025MACEOFF} models.
     The timing of MD simulations for bulk liquid water with varying system sizes was performed on an NVIDIA L40S GPU.
     }
    \label{fig:MACELES-OFF_timing}
\end{figure}

Fig.~\ref{fig:MACELES-OFF_timing} compares the inference speed of the MACE-OFF23 (S, M, L) models and the MACELES-OFF model for MD simulations of liquid water using both the ASE and OpenMM implementations on a single NVIDIA L40S GPU with 48~GB of memory. It shows that the computational overhead introduced by including long-range interactions is minimal. 
The MACELES-OFF model shows a comparable speed to the MACE-OFF23(S), which shares the same cutoff radius ($r=4.5$~\AA{}), and is significantly faster than the MACE-OFF23(M) and 23(L). There is no significant speed difference between ASE and OpenMM for any of the models. Moreover, all models demonstrate favorable scaling. The MACELES-OFF model supports simulations with up to approximately 6,000~atoms on a single GPU, while the MACE-OFF23(S) model supports up to around 15,000 atoms.

\textbf{Data availability}~The training sets, training scripts, and trained potentials are available at \url{https://github.com/ChengUCB/les_fit}.

\textbf{Code availability}~
The LES library is publicly available at 
\url{https://github.com/ChengUCB/les}.
The CACE package with the LES implementation is available at \url{https://github.com/BingqingCheng/cace}.
The MACE package with the LES implementation is available at \url{https://github.com/ChengUCB/mace}.
The NequIP package with the LES implementation is available at \url{https://github.com/ChengUCB/nequip}.
The MatGL package with the LES implementation is available at \url{https://github.com/ChengUCB/matgl}.
% The CACE, MACE, NequIP, and MatGL packages with LES implementation are publicly available at \url{https://github.com/BingqingCheng/cace},
% \url{https://github.com/ChengUCB/mace},
% \url{https://github.com/ChengUCB/nequip},
% and \url{https://github.com/ChengUCB/matgl}, respectively.
% The long-range method is implemented as an \texttt{Ewald} module. 

\textbf{Acknowledgements}~
D.K. and B.C. acknowledge funding from Toyota Research Institute Synthesis Advanced Research Challenge. T.J.I, D.S.K. and P.Z. acknowledge funding from BIDMaP Postdoctoral Fellowship. 
The authors thank Bowen Deng for valuable discussions on MatGL implementation,
and thank Gabor Csanyi for stimulating discussions.

\textbf{Competing Interests}
B.C. has an equity stake in AIMATX Inc.
The University of California, Berkeley has filed a provisional patent for the Latent Ewald Summation algorithm.

%\printbibliography

\end{document}